\documentclass[preprint]{aastex63} 

\usepackage{graphicx}
\usepackage{amsmath,amssymb}

\shorttitle{Time spectroscopy of Epsilon~CrA}
\shortauthors{Rucinski}

\begin{document}

\title{Time sequence spectroscopy of Epsilon~CrA.\\
The 518 nm Mg~I triplet region analyzed with
Broadening Functions
}

\author{Slavek M. Rucinski} 
\affil{Department of Astronomy and Astrophysics, 
University of Toronto \\
50 St. George St., Toronto, Ontario, M5S~3H4, Canada}
\email{rucinski@astro.utoronto.ca}

\begin{abstract}
High-resolution spectroscopic observations of
the W~UMa-type binary $\epsilon\,$CrA obtained as a 
time monitoring sequence on four full and 
four partial nights within two weeks have been used to 
derive orbital elements of the system and discuss
the validity of the Lucy model for description of the radial-velocity data. 
The observations had more extensive temporal coverage and
better quality than similar time-sequence observations 
of the contact binary AW~UMa. 
The two binaries share several physical properties with both showing 
very similar deviations from the Lucy model: The
primary component is a rapidly-rotating star  
almost unaffected by the presence of the secondary component, while
the latter is embedded in a complex gas flow and appears to have its own 
rotation-velocity field, in contradiction to the model. 
The spectroscopic mass ratio is found to be larger 
than the one derived from the light-curve analysis, similarly as in many other
W~UMa-type binaries, but the discrepancy for $\epsilon\,$CrA 
is relatively minor, much smaller than for AW~UMa. 
The presence of the complex velocity flows contradicting the
solid-body rotation assumption suggest a necessity of modification
to the Lucy model, possibly along the lines outlined by 
\citet{Step2009} in his concept of the energy transfer between the 
binary components. 
\end{abstract}
    

\section{Introduction}
\label{intro}

$\epsilon\,$Coronae Austrinae (Epsilon~CrA, hereinafter $\epsilon\,$CrA; 
HD~175813, HR~7152, F2V, $V = 4.8$, $B-V=0.36$, $P=0.591$ d.)  
is the brightest known binary star of the W~UMa type. Such stars 
are also frequently called ``contact binaries'' because they achieve 
a state of physical equilibrium in actual physical contact 
permitting free exchange of energy between their components. 
The result is the observed equality of the effective temperature over 
the whole binary structure.
Since the component masses always differ, with typical mass ratios
around 0.3 to 0.5 and the steep mass-luminosity relations, 
the more massive component provides practically all
luminous energy while the less massive 
component carries the angular momentum 
of the binary and contributes its radiation area 
to the combined system luminosity. 
The first physically acceptable structure model of the W~UMa binaries
was developed by \citet{Lucy1968a}. The companion study,
\citet{Lucy1968b}, used its assumptions to give a
prescription on light-curve models utilizing the Roche-model
equipotentials. While the light-curve model
enjoys immense popularity with hundreds or thousands of
light curves synthesized so far, the large body of observational 
results has not resulted in much progress in the understanding of the 
energy exchange process which is so crucial for the structure model.
A vigorous and often heated discussion of the 1970's 
died out when it was realized that most severe controversies
cannot be resolved using physical arguments without observational 
support. As a result, the structure model of \citet{Lucy1968a} is
assumed valid irrespective of the energy transfer processes 
operating in these binaries. 
We return to this important subject at the end of the paper,
 in Section~\ref{new}, in a discussion of structure and energy 
 transport models.

The current study has been motivated by the similarity 
$\epsilon\,$CrA to AW~UMa, the first and currently the only contact 
binary which has been studied spectroscopically 
in sufficient detail using the  high-resolution, continuous,
time-monitoring observations \citep[Paper~I]{Rci2015}. 
Both binaries have small mass ratios close to 0.1. It is
not clear if rather moderate numbers of known such small 
mass-ratio W~UMa binaries is an observational selection bias 
or a genuine property 
of the group;  extensive sky surveys will soon resolve this matter.
We assume that both binaries are contact binaries with the same
processes operating in them as in more common systems
with mass ratios around 0.3 -- 0.5.
The results for AW~UMa were surprising and in discord with the
Lucy model: the binary turned out to be a puzzling case of
a binary which observed photometrically --  and analyzed for the light-curve variations
-- seemed to be a perfect illustration of the Lucy model, and yet -- spectroscopically 
-- appeared to resemble a semi-detached binary with the primary losing 
its matter and engulfing the secondary in its outer layers. 
The current observations and analysis of  $\epsilon\,$CrA were meant as
an attempt to answer the question: Is AW~UMa an atypical binary? Or,
more importantly, is the dichotomy in the photometry versus spectroscopy telling us 
that photometry does not carry sufficient information through its
spectrum-integrating nature? In other words, is bringing one more dimension into
the temporal analysis necessary to see the full complexity of such a binary system?

Radial velocities are  easy to predict with the Lucy model \citep{Rci2012}. 
The model envisages 
the two stellar components filling the same common Roche-model 
equipotential and rotating as a solid body with a total 
absence of velocities in the rotating system of coordinates.
Extensive use of the model for the moderate spectral resolution
observations of  90 W~UMa binaries
at the David Dunlap Observatory (a very brief summary is in
 \citet[Sec.~2]{Rci2013})  appeared to confirm the model. 
However, the high-resolution, high-quality data presented in Paper~I 
showed that AW~UMa contradicts the Lucy model and indicated a need for
its modifications. The essential results of Paper~I (see also the earlier
work on AW~UMa by \citet{PR2008}) are as follows:\\
1.  The upper 70\% of the rotational profile of 
the rapidly-rotating primary of AW~UMa
agrees with that of a single star unaffected by the presence of its companion
and rotating at the projected equatorial velocity
$V \sin i$ of about 85\% of the orbital synchronism. \\
2.  An additional, high rotation-rate velocity component called 
``the pedestal'' extends the profile width to that expected for 
the rotating Roche-model structure. \\
3.  The surface of the primary component is covered by more regularly distributed  
``ripples'' and relatively isolated ``spots''. These features may be 
located at different stellar latitudes with different rotation rates, as indicated 
by different rates of migration across the stellar profile.\\
4. The secondary component possesses a velocity field  embedded in the
inter-binary gas flow. Its flat or meniscus-shaped rotation profile 
does not show a central condensation, suggesting 
a detached star of unusual velocity distribution rather than a 
component of the Lucy model.\\
5. In spite of the velocity field modified by the inter-binary flows,
the phase dependence of the integrated spectral-line intensities
(integrated equivalent widths) (Paper~I, Fig.~2) roughly agrees
with the photometric data.
The variations 
can be explained by making a simple assumption of a phase-invariable primary
with all equivalent-line-width phase variations due to changing 
visibility of the secondary component and of its surrounding matter.\\
6. The masses of the two components of AW~UMa 
are: $M_1 \sin^3 i = 1.29 \pm0.15\, M_\odot$, 
$M_2 \sin^3 i = 0.128 \pm0.016\, M_\odot$ (for the assumed $i \simeq 80^0$
 as for the Lucy model). Irrespectively of the model, the orbital inclination is not far
from the edge-on orientation so that the above minimum-mass estimates
are close to the actual ones. 
The spectroscopic mass ratio, $q_{\rm sp} = 0.10$, 
is different from the indicated by photometric results using 
the Lucy model, $q_{\rm ph} = 0.08$, with the 
formal uncertainties of both methods several tens-of-times 
smaller than their difference. \\
7.  The moderate period change of AW~UMa of $dP/dt = -5.3 \times 10^{-10}$
\citep{Rci2013a} could be interpreted by a net mass-transfer 
from the primary to the  secondary at a rate of about 
$dM/dt  \simeq -2 \times 10^{-8} M_\odot/{\rm yr}$. 

Paper~I, although presenting unexpected results for AW~UMa, 
has gone critically uncommented except for the important 
investigation by \citet{Eaton2016} who carefully 
considered possible ways of reconciling
the new spectroscopic results with the Lucy light curve model. Several of the
suggested modifications (e.g.\ large photospheric spots) would imply the
introduction of many arbitrary parameters just to fit the data,
but some criticism may be valid
(e.g.\  the use of the flux spectrum as a template in view of the complex 
geometry of the local light emergence). 

The observations of $\epsilon\,$CrA reported in this paper required  extensive
use of a very stable, high-resolution spectrograph on a moderate-size telescope.
Such observations are not easy to arrange and execute. 
Because of the high brightness of the star and several and some organizational reasons, 
it turned out easier to arrange a continuous spectral time-monitoring 
for $\epsilon\,$CrA than for V566~Oph for which we had important indications 
from lower-resolution work \citep{PR2008,Rci2010} that the Lucy 
model may work correctly. Thus, instead of  attempting to analyze an object
entirely different from AW~UMa, we concentrated on the uniqueness 
of that binary and on general lessons for the most 
appropriate model of W~UMa binaries. 

Although $\epsilon\,$CrA with $V = 4.8$ is a relatively bright star, 
the previous analyses on $\epsilon\,$CrA have not been extensive. The early 
light-curve studies of \citet{Knip1967} and \citet{Tapia1969} have
remained the main photometric source for the Lucy model light-curve studies of 
\citet{Tw1979} and, especially \citet{WR2011}, who pushed the Lucy model 
to its full, extensive consequences with many parameters determined 
at the same time. The orbital inclinations were determined
to be $i = 72.24 \pm0.50$ degrees \citep{Tw1979} and 
$i = 73.05 \pm0.16$ degrees  \citep{WR2011}, both consistently smaller than the
orbital inclination of AW~UMa estimated at $i \simeq 78 - 80$ degrees
(see Paper~I). 
The photographic line-by-line radial-velocity analysis by \citet{TW1975}
was substantially improved by the study of \citet{GD1993} who detected
the secondary component using the cross-correlation function (CCF) technique,
finding the mass ratio of $q_{\rm sp} = 0.128 \pm0.014$. This mass ratio is 
significantly larger than the formally very accurate photometric results of 
\citet{Tw1979}, $q_{\rm ph} = 0.114 \pm0.003$ 
and  \citet{WR2011}, $q_{\rm ph} = 0.1244 \pm0.0014$. 
The study by \citet{GD1993} was the closest in spirit to the current one as our 
Broadening-Functions technique can be considered an 
improvement on the CCF,  but  based on a linear transformation 
rather than a non-linear one \citep{Rci2002} and retaining the
full spectral resolution of the data. 
The reader is advised to consult Paper~I as the 
current paper is very closely related to the AW~UMa study \citep{Rci2015}.

The $\epsilon\,$CrA spatial velocity does not differ appreciably 
in its general properties from the majority of the nearby W~UMa binaries 
\citep{Rci2013} which appear to be galactic-disk, kinematically-old objects.
With the assumed: $\mu_{\rm RA} = -132.2 \pm0.41$ mas/yr,
$\mu_{\rm Dec} = -98.4 \pm0.38$ mas/yr and $\pi = 32.01 \pm0.27$ mas 
\citep{Gaia2016,Gaia2018}                                        
and our new $V_0 = 62.541$  km~s$^{-1}$ 
(Section~\ref{sec} and Table~\ref{tab_els}),  
the spatial velocity components are
$U=+71.9$, $V=-7.8$ and $W=+0.2$, all in km~s$^{-1}$, with the convention 
on the $U$ vector direction as in \citet{Rci2013}.
The tangential velocities are moderate and it is only the radial direction
 in which the space-velocity vector is larger. 
The absolute magnitude $M_V = +2.3$ implies a relatively
bright star for its spectral type. In addition
to the most frequently cited F2V, the spectral type of $\epsilon\,$CrA 
was estimated by  \citet{Grey2006} to be as late as F4VFe-0.8. 
Both, the high luminosity and the slightly later spectral type than of AW~UMa
suggest a binary object with stellar cores  somewhat advanced in their evolution.

The paper presents the observations of $\epsilon\,$CrA in Section~\ref{obs}
and their results in Section~\ref{res}. The new results and those for
AW~UMa suggest a re-discussion of the venerable Lucy model. The
essential directions for such an effort in view of the results for both binaries
are outlined in Conclusions in Section~\ref{mod}.


\section{Observations and their processing}
\label{obs}

Observations of $\epsilon\,$CrA were made in the service mode
with the CHIRON spectrograph \citep{Toko2013} 
on the CTIO/SMARTS 1.5m telescope on 
eight nights between 12 to 29 July 2018. Six nights were within the first week while
the remaining two were two weeks later (see Table~\ref{tab_obs}).
The first three nights were clear, but the following nights were affected 
to a various degree by the variable
weather conditions of the Chilean winter. 
The overall span of the $\epsilon\,$CrA 
observations of 17 days is much longer than for AW~UMa for which 
all observations were collected in three consecutive nights.
Altogether, 
361 exposures of 450 seconds were taken using the slicer setup which provides 
a spectral resolving power of $R = 80,000$. An attempt was made to retain the same
time resolution throughout the program. The median spacing between
the observations was 468 seconds or 7.8 minutes which is a less rapid
data-taking rate than used
for AW~UMa where the median cadence was 2.1 minutes\footnote{The
AW~UMa observations were done in the polarimetric mode with four consecutive 
spectra forming one measurement. These spectra were utilized in Paper~I individually 
disregarding the polarization signal 
after no measurable polarization was detected in the data.}. 
The longer exposures and more observation nights resulted in the total
amount of exposure time for $\epsilon\,$CrA of 2705 minutes; this should be compared with
876 minutes for AW~UMa. 

\begin{figure}[h]
\begin{center}
\includegraphics[width=14.0cm]{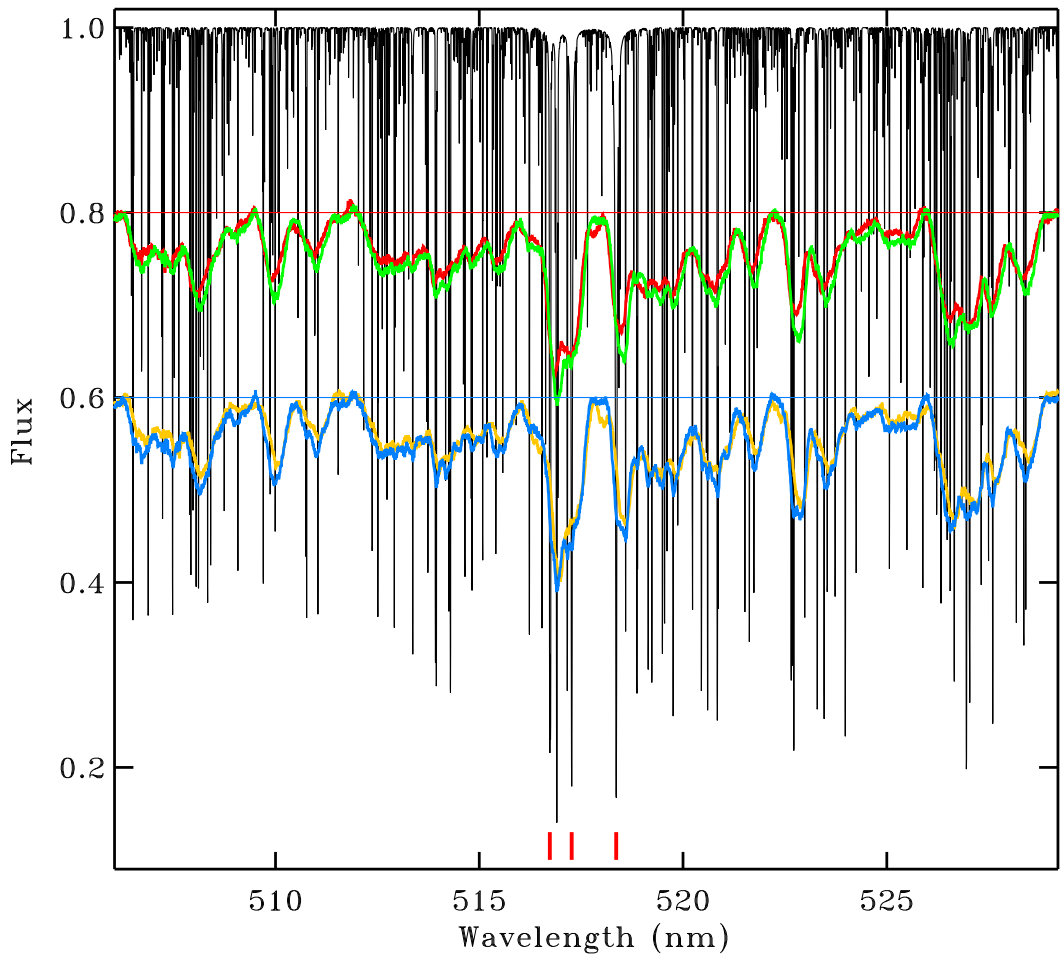}    
\caption{
\footnotesize 
The model-spectrum template used for the Broadening-Functions deconvolution
technique (thin black line) and the observed spectra of
$\epsilon\,$CrA (in color) are shown together, with the observed spectra shifted
by $-0.2$ for the orbital phases 0.25 and 0.5 (red and green lines) 
and $-0.4$ for the phases 0.75 and 0.0 (orange and blue lines), respectively. 
The spectral region is centered on the Mg~I ``b'' triplet. 
The integrated spectral-line absorption in the whole spectral window is 5.51 percent
and is due mostly to iron lines with the magnesium triplet contributing  0.64 percent 
to that number. The lines of the  triplet are marked by the red bars 
at the bottom part of the figure. 
}
\label{fig_spec}
\end{center}
\end{figure}


\begin{deluxetable}{cccc}

\tabletypesize{\footnotesize}    
\tablewidth{0pt}
\tablecaption{Observing nights of $\epsilon\,$CrA
     \label{tab_obs}}

\tablehead{\colhead{July 2018} & \colhead{$HJD$} & \colhead{$Orb$} & \colhead{$N$} 
}

\startdata
  12  &  58312.5211 --  58312.8785  &  0.7601 --  1.3643  &  66 \\
  13  &  58313.5402 --  58313.8870  &  2.4830 --  3.0693  &  64 \\
  14  &  58314.5356 --  58314.8600  &  4.1660 --  4.7145  &  60 \\
  15  &  58315.5380 --  58315.5631  &  5.8608 --  5.9033  &  6 \\
  16  &  58316.5378 --  58316.8031  &  7.5513 --  7.9998  & 49 \\
  18  &  58318.6736 --  58318.8660  &  11.1622 -- 11.4876 &  36 \\
  26  &  58326.5976 --  58326.7576  &  24.5597 -- 24.8304 &  30 \\
  28  &  58328.5632  -- 58328.8552  &  27.8830 -- 28.3768 &  54
\enddata

\tablecomments{The mid-exposure time ranges for individual nights  are
expressed as $HJD - 2,400,000$.  $Orb$ gives the range
in time in units of the orbital period, continuously counted from the assumed 
initial epoch $HJD = 2\,458\,312.0716$ and $P = 0.59145447$ day; the
fractional part of $Orb$ is the  orbital phase. 
$N$ is the number of star spectra obtained on a given night.
The nights are  identified by the day in July 2018.}

\end{deluxetable}

The spectral region used for $\epsilon\,$CrA was 506.05 nm to 529.20 nm, 
which is slightly wider at the short-wavelengths than the region adopted for AW~UMa
additionally contributing to improved final results.  
The spectral sampling at the conversion step from wavelengths to velocities was
set  at 1.8 km~s$^{-1}$ leading to 7450 points in the utilized spectrum.
Four typical spectra for the cardinal orbital phases of the binary are shown in 
Figure~\ref{fig_spec} together with the spectrum of the template used 
for the Broadening-Functions (BF) processing. The 
spectra are dominated by the strong rotational broadening at all orbital phases. 
The orbital-phase variations -- the subject of this paper --
are rather subtle, mostly masked by strong overlapping of the 
broadened lines. This stresses the necessity of high quality data and of an
efficient information-extraction technique. The Broadening-Functions
technique in the most convenient as it derives 
information on the velocity field using a linear
deconvolution of the rotation- and orbital motion-broadened spectrum. 
The final product is a function similar to a cross-correlation, but having an
important advantage of  linearity and thus of a direct mapping of the radial 
velocity field into the spectral wavelength domain. The function integral
is normalized to unity for a perfect match of the integrated equivalent width of
the spectral lines to that of the template spectrum. The function recovers 
the actual level of the spectral continuum which is not accessible in the
wavelength domain.

Parameters of the BF processing
of the $\epsilon\,$CrA data were set at the same values as for the case 
of AW~UMa, as described in Paper~I. 
Similarly, the spectra were smoothed to the effective 
resolution characterized by a Gaussian with the FWHM of 
8.5 km~s$^{-1}$, corresponding to the resolving power $R \simeq 35,000$.
The Broadening-Functions (BF) processing was used to determine 
the functions at 551 points leading to the strong, 13.5-fold 
over-determinacy of the final, individual profiles. 
As the template, we used the same model-atmosphere flux spectrum 
of a F2V star, calculated by Dr.\ Jano Budaj for AW~UMa. 
Because of the very wide spectral lines and the atypical appearance of the
spectrum, the spectral type of $\epsilon\,$CrA has been estimated in the literature
as ranging between F0V and F4V; the color index $B-V = 0.36$ agrees with F2V
(for AW~UMa: $B-V = 0.33$). 
The BF's calculated for the available spectra of $\epsilon\,$CrA are
characterized by a median $S/N \simeq 120$. Thus, they are substantially 
more accurate than those for AW~UMa ($S/N \simeq 57$) reflecting the longer 
exposures and the higher apparent brightness of $\epsilon\,$CrA.
The BF's are listed in Table~\ref{tab_BFs}. The overall size of the table
is almost $2\!\times\!10^5$ numbers stored in a compact, integer form;
storage of the  spectra would take 13.5 times more space\footnote{The raw 
spectra are available from the CTIO-CHIRON archive site, 
http://www.ctio.noao.edu}.


A prediction of the moment of primary (deeper) eclipses of the more-massive
star by the secondary component, based on photometric data have been
provided to the author by Prof.\ J.\ Kreiner and Dr.\ B.\ Zakrzewski of Cracow.
The data covered the years 1960 to 2017 so that the ephemeris required extrapolation
forward in time. The quadratic ephemeris is: 
$HJD = 2\,439\,707.6624(2) + 0.59143046(23) \times E + 3.817(79)\,10^{-10} \times E^2$,
with the uncertainties of the last digits given in parentheses.
For an eclipse occurring just before the start of our observations, the epoch and time are:
$E=41456$,  $HJD = 2\,458\,312.0766$. 
A sine fit to the radial velocities of the primary component of $\epsilon\,$CrA 
(see Section~\ref{pri}) ) gave a slightly different value 
for the mid-primary eclipse which has been adopted
as the origin for our orbital-phase reckoning. The final, locally-linear
prediction for the primary eclipses that we used was: 
$HJD = 2\,458\,312.0716 + 0.59145447 \times E_1$, with $E_1=1$ for
the first of the observed eclipses. 

A continuous measure of time, but expressed in the orbital period as a unit, is
a convenient argument to see changes taking place in AW~UMa over a few orbital cycles.
We use the same quantity for $\epsilon\,$CrA, equivalent to $E_1$ above, 
but expressed as a real number with its fractional part being the familiar orbital phase
called hereinafter $\phi$; we call it the ``orbit'' ($Orb$). Our data covered the orbits 
$0.760 \le Orb \le 28.377$.

\begin{figure}[h]
\begin{center}
\includegraphics[width=12.0cm]{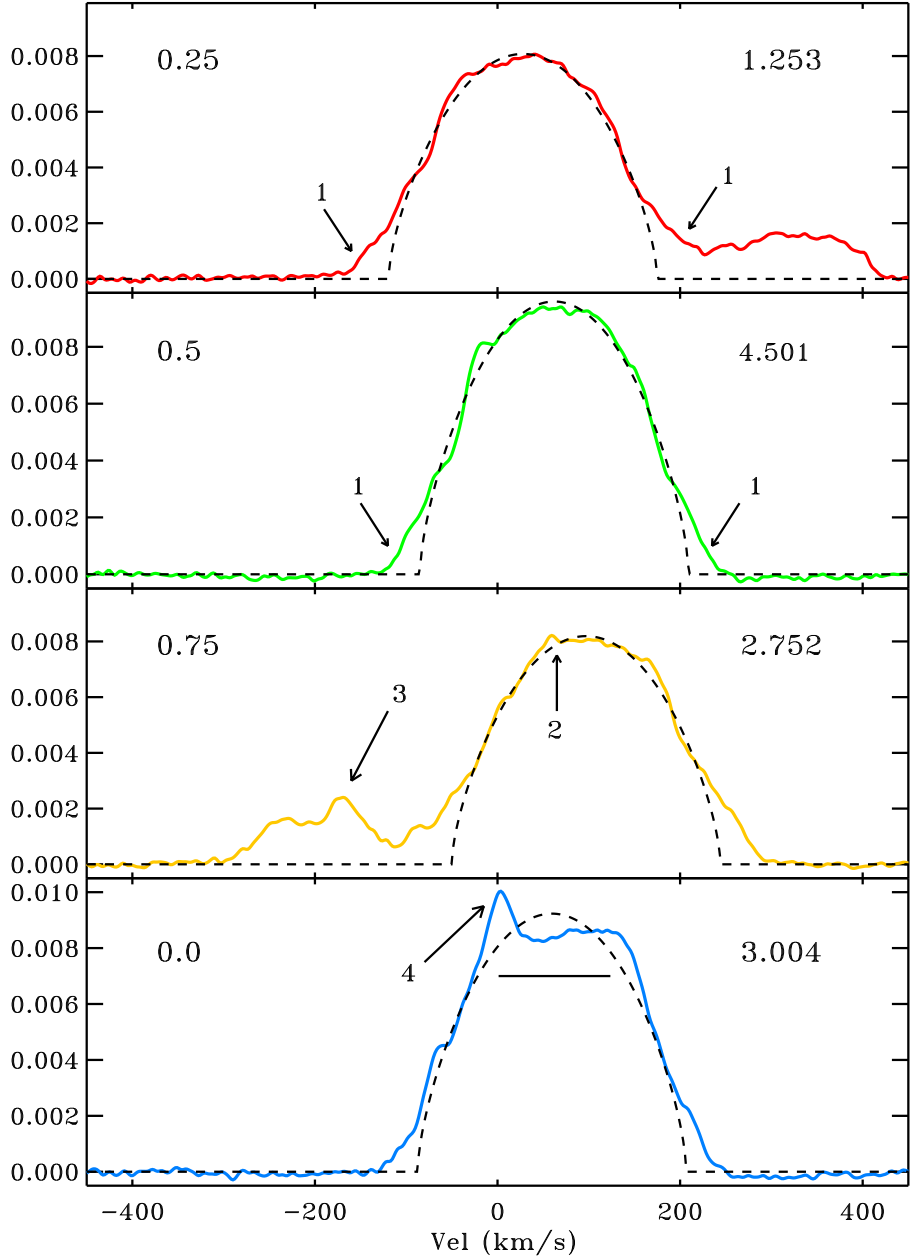}    
\caption{
\footnotesize 
Broadening Functions (BF) for four observations close to  
 orbital quadratures and mid-eclipses with the color coding corresponding
 to that used in Figure~\ref{fig_spec}.
The nominal orbital phases are given in the left upper corners of the panels
while the orbits ($Orb$, see the text) of the actual observations are in the right
upper corners. The units of the vertical scale relate to the
normalization of the BF integrals to the template when sampled at 1.8 km~s$^{-1}$
(unity for the exact spectral match of the spectra). 
The BF features indicated by the 
numbers are: 1 -- the pedestal; 2 -- a surface inhomogeneity;
3 -- the ``wisp'', a radial-velocity structure associated with the secondary 
component of the binary (Section~\ref{stream});
4 -- the flow of matter over the surface of 
the secondary component directed toward the observer (Section~\ref{pri}). 
The broken lines give the vertically-scaled rotational profile for 
$V \sin i = 147$ km~s$^{-1}$. In the lowest panel, the horizontal bar
shows the expected velocity extent ($\pm64$ km~s$^{-1}$) 
of a Lucy-model, synchronously-rotating secondary component 
with $q=0.13$, for $i=73$ degrees; the secondary is 
filling the inner critical common equipotential passing
through the $L_1$ point, see Section~\ref{sec} and Table~\ref{tab_RV}. 
}
\label{fig_4ph}
\end{center}
\end{figure}


\section{The results}
\label{res}

\subsection{The primary component}
\label{pri}

The Broadening Functions (BFs) for  $\epsilon\,$CrA 
are very similar to those of AW~UMa in that they are dominated
by the strong feature of the primary component of the binary system.  
We show the BFs for four cardinal orbital phases in 
Figure~\ref{fig_4ph}; the results for other observations
are tabulated in Table~\ref{tab_BFs}. We see that:
(1)~The rotational profile of single, rapidly rotating star with  
$V \sin i = 147$ km~s$^{-1}$ 
describes the upper part of the primary component profile (above 
the 0.35 of the maximum) very well; the transit eclipse being an obvious exception; 
(2)~The profile shows high-velocity 
extension wings which we called  the ``pedestal'' for the case of AW~UMa; 
(3)~The primary component surface appears to be covered by low-contrast
inhomogeneities to a lesser degree than in AW~UMa, and 
(4)~Obvious signs of distributed matter in the binary are present and 
may be affecting the primary profile. One of such is a conspicuous BF 
perturbation at the primary (deeper) eclipse during the transit of
the secondary over the disk of the primary 
component (the lowest panel in Figure~\ref{fig_4ph}). The perturbation,
in the form of a spike is 
most likely caused by gaseous matter flowing towards the observer 
over the surface of the secondary component; this is discussed 
further in Section~\ref{sec}.


\begin{deluxetable}{RRRRRRRR}

\tabletypesize{\footnotesize}          
\tablewidth{0pt}
\tablecaption{Broadening Functions 
   \label{tab_BFs}}

\tablehead{ &  &  &  &  &  &  &  
}

\startdata
            &  -4950  &  -4932  &  -4914  &  -4896  &  -4878  &  -4860  & ... \\ 
  5211  & -336     & -159     & -39       &   19      &          9  &    -46    & ... \\      
  5263  & -330     & -202     & -137     &   -99     &      -63  &     -25   & ... \\
  5319  & -289     & -113     &   -4       &   41       &      27   &     -32   & ... \\
  5372  & -419     & -256     &  -145    &  -64      &       -8   &      19    & ... \\
  5426  & -313     & -167     &   -80     &   -29      &      -3   &       -6    & ... \\             
   ...     &              &             &             &              &             &              &  
\enddata

\tablecomments{
The table has 361 rows of 551 numbers corresponding to the number
of observations and the length of the BF functions. 
For efficient transfer, the ASCII file has
all numbers scaled and converted to integers. 
The first row gives 551 radial velocities in the heliocentric system, 
expressed in km~s$^{-1}$ and multiplied by 10 (format: 6x,551i6). 
The subsequent 361 rows (format: 552i6) give, in the first position, 
the time $T = HJD-2\,458\,312$ multiplied by $10^4$,  and then
551 values of the Broadening Function multiplied by $10^6$.
Thus, the velocities in the first row are: 
$-495.0$, $-493.2$, $-491.4$, $-489.6$, $-487.8$, $-486.0$, ... km~s$^{-1}$
and the corresponding BF values at $T = 0.5211$ are:
$-0.000336$, $-0.000159$, $-0.000039$, $0.000019$, $0.000009$, $-0.000046$.\\
This table is available in the on-line version only. 
}

\end{deluxetable}

The details in the profiles in Figure~\ref{fig_4ph}
are all significant and are not  caused by the  BF uncertainties which are very small 
(Section~\ref{obs}). 
The errors have a constant distribution along the whole
length of a given BF; their size may be estimated directly from the scatter 
in the baseline, for example from the empty portion
at velocities $<350$  km~s$^{-1}$. It is one of the
advantages of the BF approach that all points are treated equivalently 
and independently by the linear deconvolution process. The spectral resolution 
of FWHM of 8.5 km~s$^{-1}$ was set at the start of the BF determination
and is the only source of correlation between the data points.

\begin{figure}[ht]
\begin{center}
\includegraphics[width=11cm]{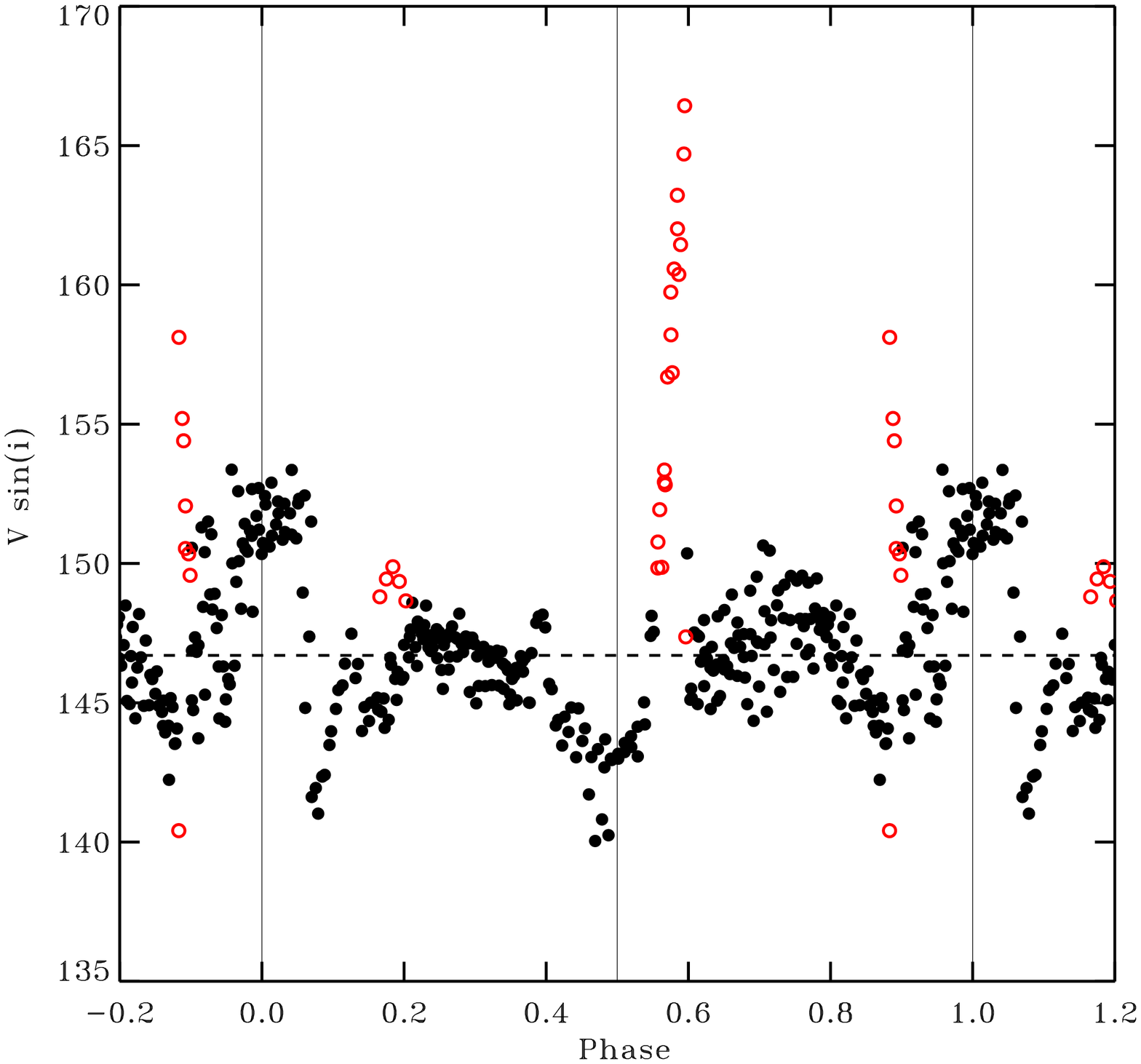}    
\caption{
\footnotesize 
The projected equatorial velocity $V \sin i$ of the primary component
of $\epsilon\,$CrA determined by rotation-profile fits to individual Broadening Functions.  
The figure shows the results of the fits to the upper 65\% of the primary-component 
profile. The data  strongly affected by the distributed matter in the system and
not used for the primary radial-velocity orbit are marked by red, open circles. 
The horizontal broken line gives the adopted value of $V \sin i = 147$ km~s$^{-1}$.
}
\label{fig_Vsini}
\end{center}
\end{figure}

\begin{figure}[h]
\begin{center}
\includegraphics[width=12.5cm]{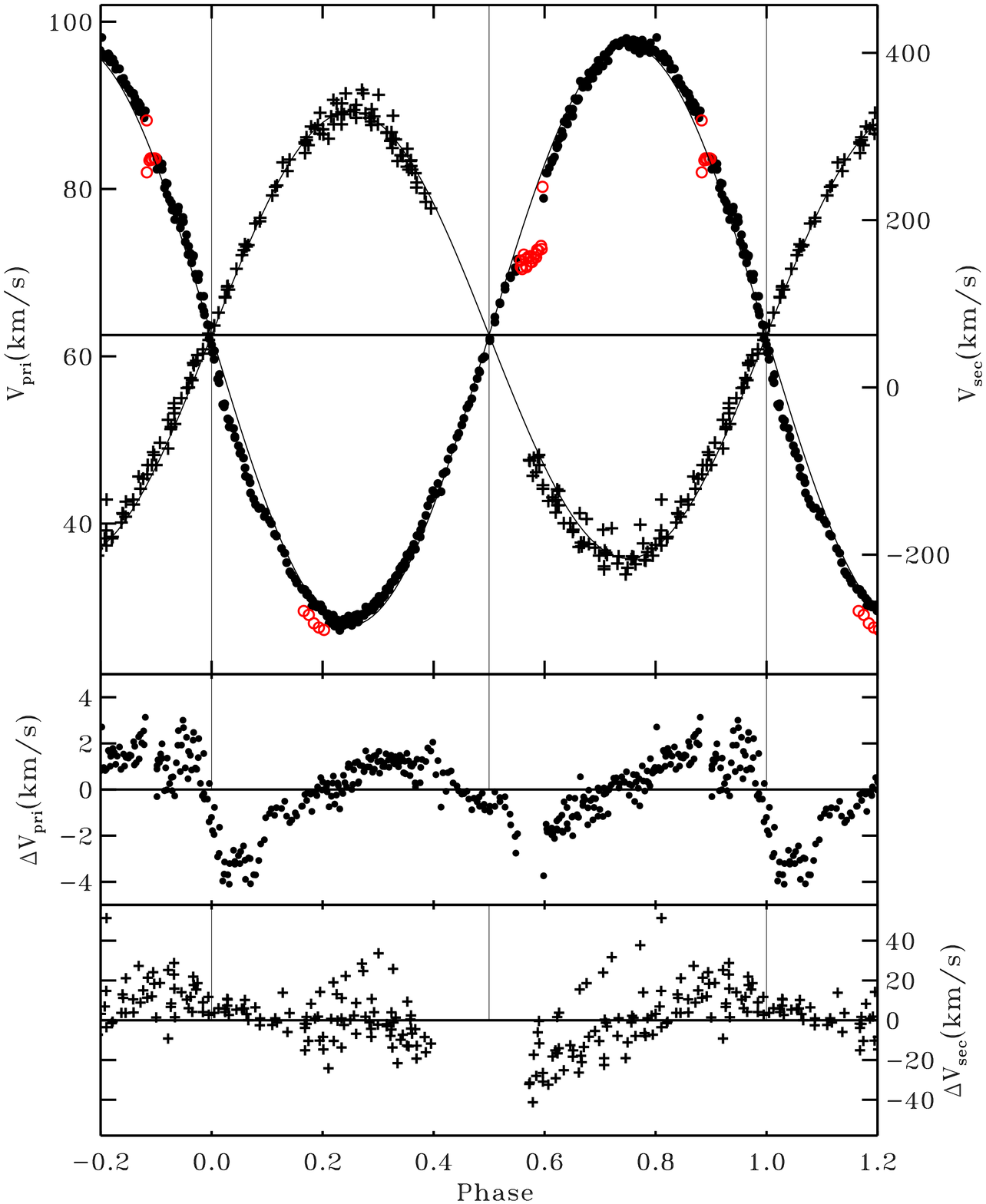}    
\caption{
\footnotesize 
The radial velocity orbit of  $\epsilon\,$CrA. The  primary component velocities 
(circles and the left vertical axis) were determined by rotational-profile fits to the upper 
65\% of the primary BF feature. The data {\it not used\/} in the sine-curve
fit are marked by red, open circles. The secondary component velocities 
(crosses, the right axis compressed by 10-times in scale relative to the left axis) were
determined as profile centroids in 2D images formed from the BF functions, 
such as Figures~\ref{night12}--\ref{night28}. The two lower panels show the velocity 
residuals, with different scales for the primary and secondary components.
}
\label{fig_RV}
\end{center}
\end{figure}

The width of the primary component profile 
(Figure~\ref{fig_Vsini}) changes slightly
with the orbital phase. The solid-body rotational profile for the limb darkening 
coefficient $u=0.5$ was fit to the individual profiles above 0.35 of the maximum 
value  to avoid contamination by the high-velocity pedestal. The fits gave 
the smallest $V \sin i  \simeq 143$ km~s$^{-1}$ during the short interval
of the secondary (shallower) eclipse, $\phi = 0.5$, 
when the secondary component with all its complications
(Sections~\ref{sec}--\ref{stream}) underwent the occultation by the primary.  
This interval appears to be longer before
the center of the eclipse, starting $\phi \simeq 0.40$ and ends abruptly 
$\phi \simeq 0.54$. These phases coincide with 
large perturbations of the primary profile by an unexplained spectral feature 
discussed further in Section~\ref{stream}.
The short duration of the totality phases agrees with the 
orbital inclination angle of $i \simeq 72-73$ degrees 
\citep{Tw1979,WR2011} estimated from the approximate relations 
of \citet{MD1972} 
and our determination of the mass ratio of $q_{\rm sp}=0.13$ (Section~\ref{sec}). 
The median projected equatorial velocity for all phases
is $V \sin i = 147 \pm3$ km~s$^{-1}$, where the assigned uncertainty 
is an estimate of the scatter at orbital phases away from the 
extended wings of the eclipses and 
not perturbed by the gas streams (Figure~\ref{fig_4ph}).
Of the total of 361 observations, 32 rotational-profile fits were noticeably 
disturbed by optically-thick gas motions in the system 
and were eliminated from the determination of
the primary component orbit (marked by red open circles in Figure~\ref{fig_Vsini}).
The rotation rate of $V \sin i = 147$ km~s$^{-1}$ is about 80--85\% of the rotation
synchronous with the orbital motion of the inner critical equipotential 
($178 \, \sin i$  km~s$^{-1}$); however, the additional broadening 
in the pedestal of about 25 -- 35 km~s$^{-1}$ would bring it into an approximate
agreement with the synchronous rate. 

\begin{figure}[h]
\begin{center}
\includegraphics[width=9.5cm]{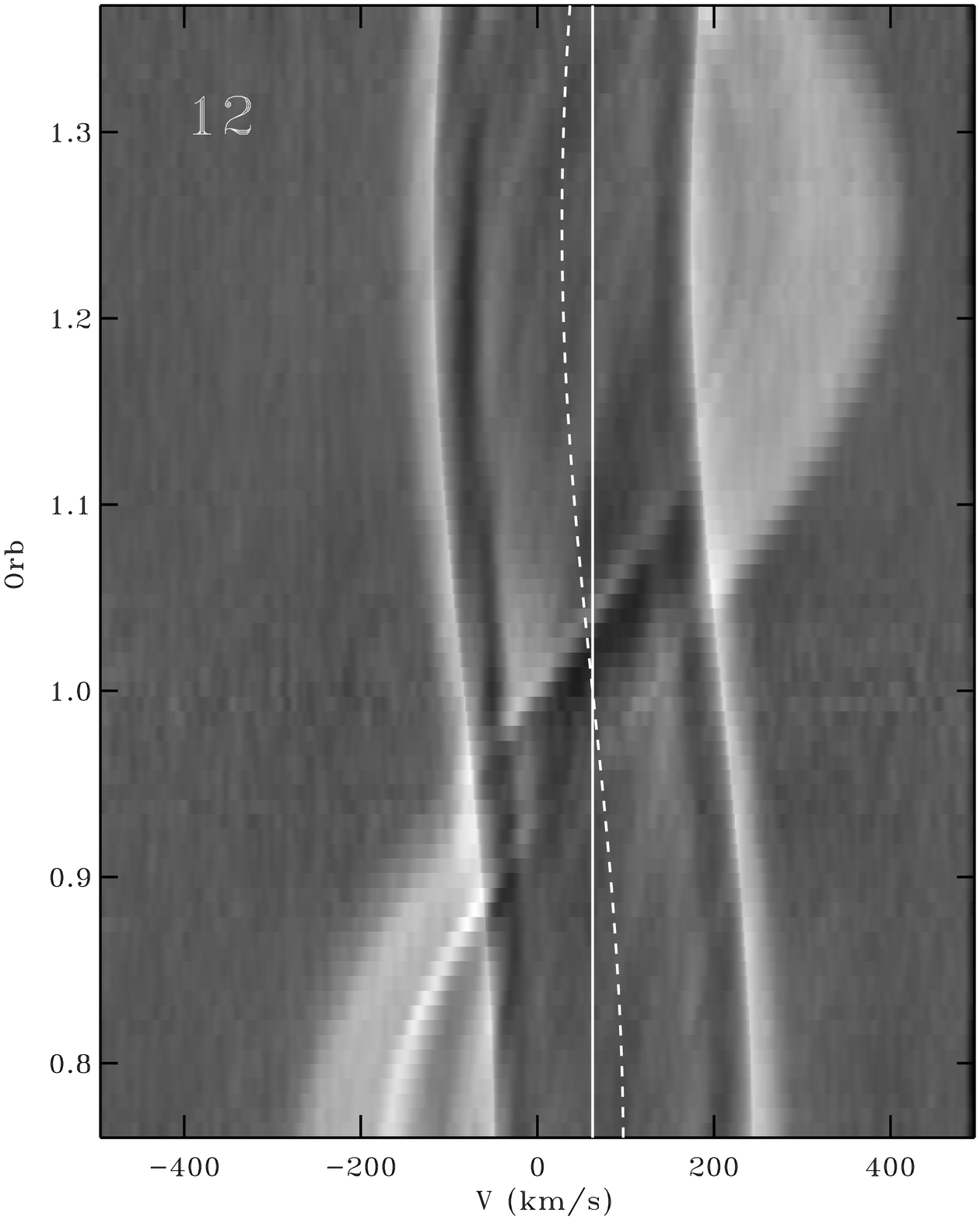}    
\caption{
\footnotesize 
The two-dimensional representation of the BF temporal variability during the
night  \#12; the nights of our program are numbered by the days of July 2018.  
The scaled and orbital-velocity shifted rotational-profile fits to the primary component 
profile, with the assumed $V \sin i = 147$ km~s$^{-1}$,  have been subtracted 
to reduce the dynamic range of the image. 
The horizontal scale gives the heliocentric velocity
while the vertical scale running upwards gives 
the time expressed in $Orb$ units; the fractional
part of the $Orb$ is the familiar orbital phase counted from the deeper (transit)
eclipse. The vertical white line gives the 
velocity of the mass center while the broken line gives 
the fit to the centroid of the primary component. 
The two bright bands along the residual profile of the primary component
show the pedestal, a feature marked in Figure~\ref{fig_4ph} by the label~1.
The enhanced, approaching side of the secondary component
is interpreted as a stream of matter over its surface 
during the transit eclipse ($orb = 1$). 
%
}
\label{night12}
\end{center}
\end{figure}


\begin{deluxetable}{ccccc}

\tabletypesize{\footnotesize}                 
\tablewidth{0pt}
\tablecaption{The radial velocities of the primary component of $\epsilon\,$CrA
\label{tab_pri}}

\tablehead{
\colhead{$HJD$} & \colhead{$Orb$} & \colhead{$V_1$} & \colhead{$V\, \sin i$}
    & \colhead{Key}
}
\startdata
     0.5211    &  0.760    &   97.84    &   149.6   &  1 \\
     0.5263    &  0.769    &   97.07    &  149.3    &  1 \\
     0.5319    &  0.778    &   97.07    &   148.4   &  1 \\
     0.5372    &  0.787    &   96.59    &   147.9   &  1 \\
     0.5426    &  0.796    &  96.63     &   147.8   &  1 \\
\enddata

\tablecomments{
The columns are as follows:\\
$HJD$ -- the heliocentric JD for the middle of the exposure counted from 
   $HJD =2458312$;\\
$Orb$ -- the same time expressed with the orbital period as an unit;\\
$V_1$ -- the centroid of the rotational velocity fitting profile expressed 
       in km~s$^{-1}$;\\
$V\,\sin i$ -- the projected rotational velocity determined from the upper 
     65 precent of the profile;\\
$Key$ --  zero signifies that the observation was not used for the primary-component
velocity determination.}

\end{deluxetable}

\begin{figure}[h]
\begin{center}
\includegraphics[width=9.5cm]{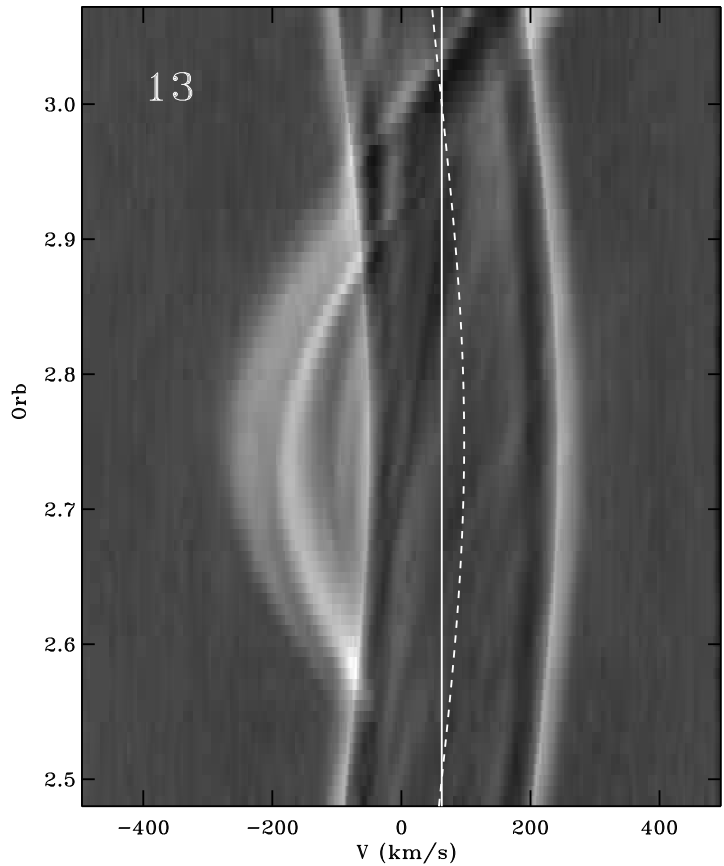}    
\caption{
\footnotesize 
The same as in Figure~\ref{night12} for 13 July 2018 when the secondary  
component was approaching us during the orbital quadrature.
Note the complex structure of the
secondary component for the orbital phases centered at $\phi \simeq 0.75$. 
The most prominent feature is called the ``wisp''; it is marked
by label 3 in Figure~\ref{fig_4ph}. In may represent a stationary spot in the
volume between the stars. 
}
\label{night13}
\end{center}
\end{figure}

The radial-velocities of the primary component of $\epsilon\,$CrA
are listed in Table~\ref{tab_pri} while the orbital solution based on the
centroids of the rotational-profile fits is shown in Figure~\ref{fig_RV}. 
The orbital fit to these velocities by a sine curve is very well defined, particularly
after the strongly deviating measurements close to the phases around 0.58 and 0.87 
are removed. The zero point determined here is used in this
paper as the origin of the orbit and orbital-phase counting.
The resulting parameters of the spectroscopic orbit for the primary
component are listed in Table~\ref{tab_els}, 
together with the results for the secondary
 component discussed in the next Section~\ref{sec}.
The orbital period $P$ is for the locally linear ephemeris
following the quadratic  elements in Section~\ref{obs}.

While the parameters of the sine-curve fit to the primary component radial velocities
have very small errors, the quality of the fit is characterized by 
a relatively large single-point random error of $\pm1.44$ km~s$^{-1}$.
As shown in Figure~\ref{fig_RV}, the large fraction of the uncertainty is due 
to the systematic deviations reaching $-4$ to $+2$ km~s$^{-1}$ close to the
eclipses; the scatter at the orbital phases around 0.25 and 0.75 is considerably smaller,
$\pm0.38$ km~s$^{-1}$. The deviations before and after the primary eclipse
(the transit of the secondary component in front of the primary)
are reminiscent of the ``Rossiter--McLaughlin effect'' effect  which was also noticeable 
in the AW~UMa  velocity data. However, observed with the accuracy available
here, the effect is not symmetric relative to the eclipse center implying 
the existence of gas-flow motions affecting even  the central parts of the 
primary profile.  

\begin{figure}[h]
\begin{center}
\includegraphics[width=9.5cm]{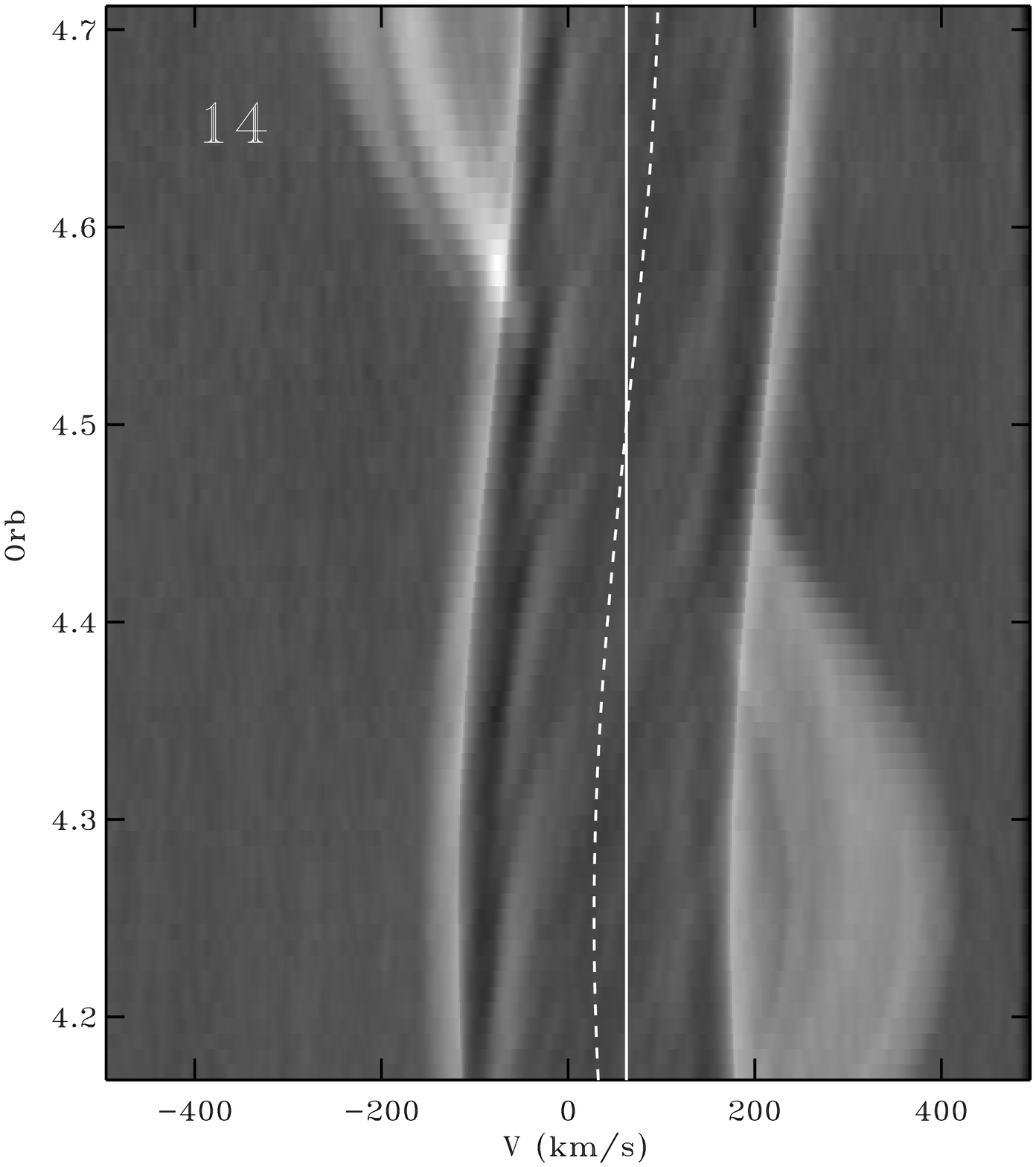}    
\caption{
\footnotesize 
The same as in Figure~\ref{night12} for the night of 14 July 2018,
centered at the shallower eclipse, the secondary-component occultation.}
\label{night14}
\end{center}
\end{figure}

\begin{figure}[h]
\begin{center}
\includegraphics[width=9.5cm]{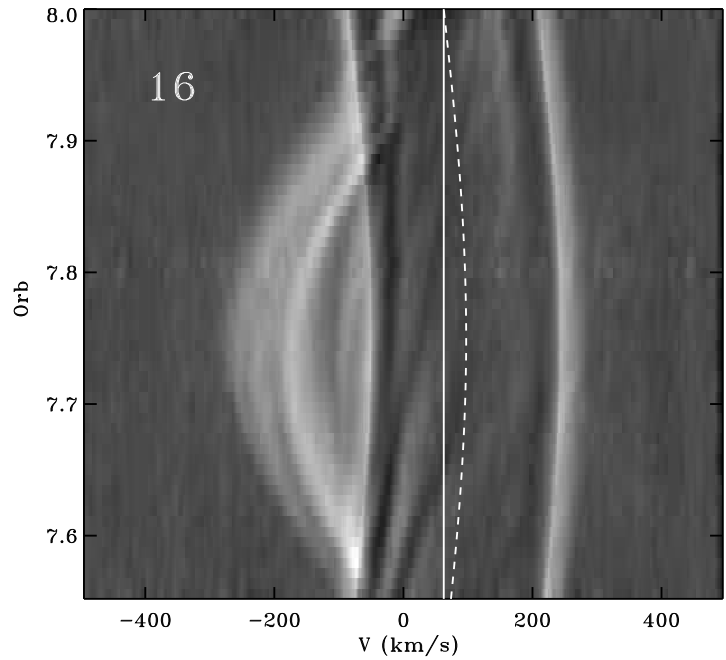}    
\caption{
\footnotesize 
The same as in Figure~\ref{night12} for the night of 16 July 2018. 
}
\label{night16}
\end{center}
\end{figure}

\begin{figure}[h]
\begin{center}
\includegraphics[width=9.5cm]{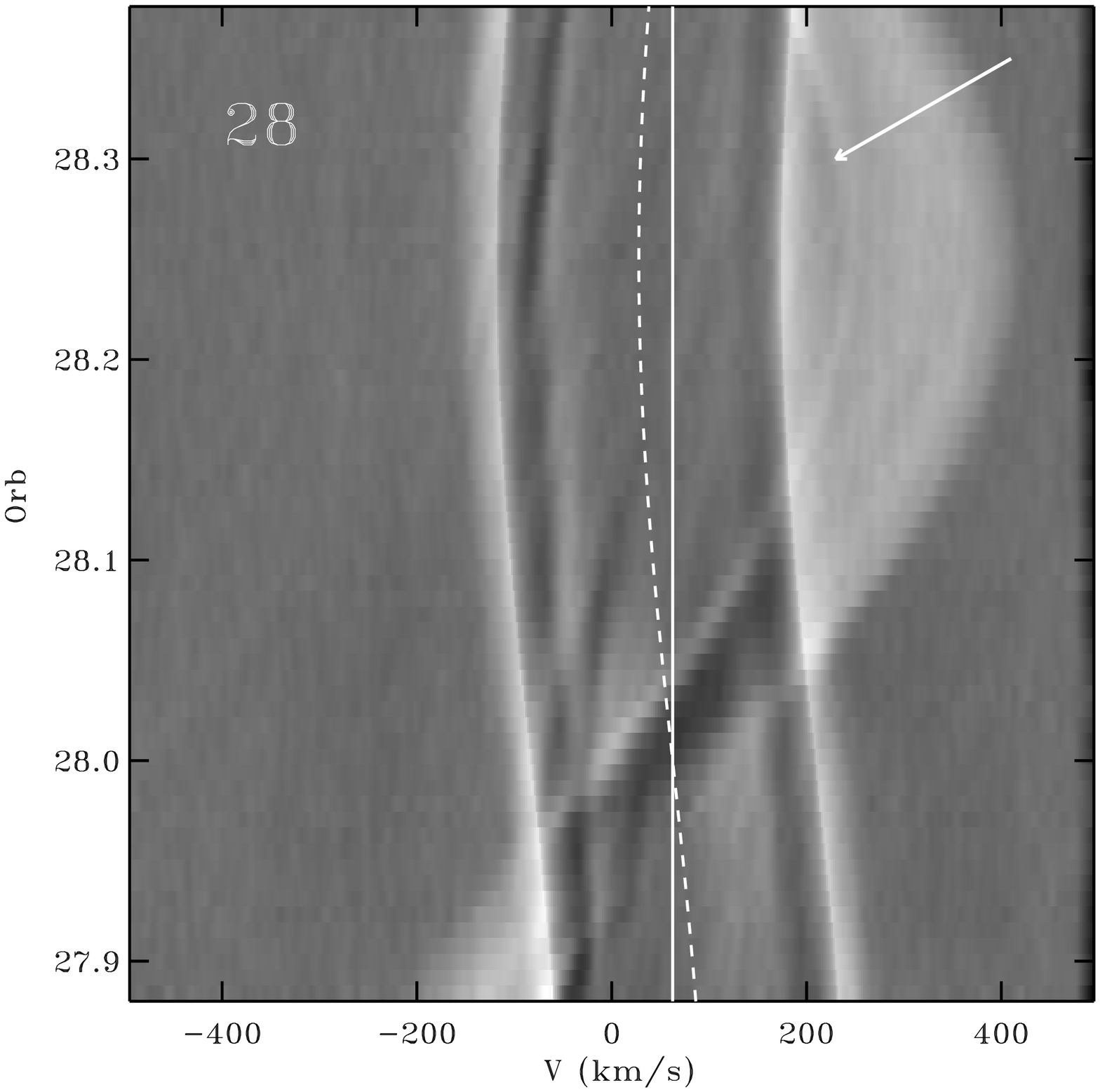}    
\caption{
\footnotesize 
The same as in Figure~\ref{night12}, but after a two-week break in 
observations, on the night of 28 July 2018.
This night included the orbital quadrature when the secondary was
receding from the observer. Note the flat profile of the secondary profile
at its positive orbital velocities and the faintly defined inner edge of the profile 
(marked by the white arrow)  indicating the velocity field of 
the secondary component detached from that of the primary component.}
\label{night28}
\end{center}
\end{figure}

Figures~\ref{night12}--\ref{night28}
show the 2D images of the BF time variability on the five nights with 
more than 40 observations of $\epsilon\,$CrA
per night (similar, smaller images exist for the three remaining nights). 
The images were obtained by interpolation into an equal-step
 grid of orbital phases with  the phase step of 0.008
 and then -- to improve the visibility of the weak features --  by subtraction of the
orbital-velocity shifted and scaled rotational-profile fits for the primary component.
Figure~\ref{Roche} should be consulted for interpretation of these figures
in terms of the sense of rotation and of the orbital phase system.  
The reader is reminded that a  connection of velocities 
to  locations and spatial dimensions is not trivial and 
requires assumptions. 

 A comparison of the 2D images of $\epsilon\,$CrA with those of
AW~UMa in Paper~I shows the primary components of
both binaries to be very similar to each other,  but with some differences: 
(1)~The large-scale disturbances visible as gentle, winding waves on the
 in the 2D pictures have smaller amplitudes in $\epsilon\,$CrA with the largest intensity
modulation of about 4\%, while in AW~UMa  some of them reached 8\%;
(2)~The small-scale ($<\!1\%$), high-frequency, regularly-spaced
 disturbances which we called ``ripples'' appear to be absent in $\epsilon\,$CrA; 
(3)~We  do not see isolated spots which may have been present on the primary
of AW~UMa. 

We conclude that the primary component of $\epsilon\,$CrA is covered by surface 
inhomogeneities to a lesser degree than AW~UMa. 
This absence of the  wavy ripples, so well visible in AW~UMa, 
is particularly significant as the BF data are of higher quality
(the $S/N$ ratio of 120 versus 57 for AW~UMa; Section~\ref{obs})
and the overall monitoring time spent on $\epsilon\,$CrA was
 3.7 times longer than for AW~UMa. However, we should remember
that the 3.1 times longer exposure time used for $\epsilon\,$CrA
may have led to some smoothing over high frequency disturbances. 
The pedestal feature of $\epsilon\,$CrA is very similar to that in AW~UMa,
remaining unexplained.

\vfill

\subsection{The secondary component and the spectroscopic orbit}
\label{sec}

In contrast to the well understood and basically phase
invariable central profile of the rapidly-rotating primary, 
the secondary component of $\epsilon\,$CrA
presents a rather complex 
picture. Its spectral lines are  weak: Expressed as integrals of the
respective BF features, the spectral lines are $7.97 \pm0.54$
times weaker than the lines of the primary component, with 
the large error in this value reflecting 
the difficulty of separating the two features.
While radial velocity profiles of the secondary can be defined in  
individual BF's, they are easier to analyze 
in the 2D images, such as  Figures~\ref{night12}--\ref{night28}
where the complexity of the profiles is fully visible  
thanks to the BF phase intercorrelation.
The images show that the secondary component
is to some extent connected to its dominating companion, 
but that the radial-velocity profiles are  different than expected
for the Lucy model.
The profiles suggest a system of gas motions or surface inhomogeneities
rather than of an unperturbed, stable photosphere.
Spatial localization of these features is generally difficult to establish, 
but this ambiguity can be sometimes lifted by the eclipses. 

A  silhouette of the secondary in radial velocities is well defined 
during the transit (primary, deeper) eclipses ($\phi = 0.0$)
when the secondary star is projected against its larger companion. 
We observed transit eclipses on four nights, numbers 12, 13, 16, and 28.
The secondary is delineated by two, symmetrically
located BF features giving its rotational
velocity half-width of $66 \pm2$ km~s$^{-1}$. 
We see a considerable difference in the appearance of the two defining ``sides'' 
of the secondary component during the transits (see 
Figures~\ref{night12}, \ref{night13} and \ref{night28}):
The negative-velocity side, at $-66$ km~s$^{-1}$, relative to the secondary 
mass center, is always more strongly defined by a
spike-like feature in the BF's.  A similar, enhanced spectral-line feature 
was observed in AW~UMa, but it is much stronger and better defined 
in $\epsilon\,$CrA (see  the  lowest panel of Figure~\ref{fig_4ph}).
We discuss this feature as a representation of inter-binary gas motions
further in Section~\ref{stream}. 

In contrast to the approaching side, the edge of the rotationally receding 
side is poorly defined  
during the transit, but its visibility improves after the primary eclipse when the
secondary continues at the positive velocities relative to the projected
mass center. These out-of-eclipse phases carry information
on the varying velocity distribution over the 
secondary surface at other aspect angles.
Of importance is the admittedly faintly-defined negative-velocity limit of the 
secondary component when this star is moving away from the observer
(the phase range $0 < \phi < 0.5$;  marked by the arrow in Figure~\ref{night28}). 
It appears that the rotational motion is at least partly 
confined to the secondary component as if in a local surface motion,
without a  connection to the velocity field of the 
primary component. The Lucy model, with its strict 
solid-body rotation, predicts no separation in velocities of 
the two components. We saw a very similar
inner edge to the secondary-component velocities 
in AW~UMa, but it is much better defined in $\epsilon\,$CrA. This implies
that in both binaries the secondary component  has its own velocity field,
 independent of the primary component.

While the secondary component of $\epsilon\,$CrA shows 
a relatively flat profile at the positive-velocity side of the primary, with 
numerous but barely-marked and tenuous structures, it is very different when it  
re-appears after the occultation at $\phi = 0.5$. 
Then, at the negative orbital velocities, it shows a more complicated
structure dominated by what appears as a ``wisp'' in the 2D images
(discussed further in the next Subsection~\ref{stream}) 
and as several delicate, phase-dependent
weaker features. The individual profiles at  negative relative velocities
are orbital-phase dependent, with only  weak traces of a 
tendency for a central-depression
 or meniscus- shaped shape, as was suggested for the less extensively
observed for AW~UMa. Generally, however, the secondary of
$\epsilon\,$CrA does not seem to show  a profile similar to that of an  
accretion disk, as was suggested in Paper~I for AW~UMa. We note
that we probably see the two binaries at orbital inclination angles
differing by about ten degrees 
so that phenomena taking place in the plane of the binary orbit may be
visible differently.


\begin{deluxetable}{cccc}

\tabletypesize{\footnotesize}                 
\tablewidth{0pt}
\tablecaption{The  spectroscopic orbit of $\epsilon\,$CrA
\label{tab_els}}

\tablehead{
\colhead{Parameter } & \colhead{Result} & \colhead{Error} & \colhead{Unit} 
}
\startdata
$P$        &   0.59145447  &                  &    day                 \\
$HJD(pri)$  &  2458312.0716  &   0.0004  &   day        \\
$K_1$    &   34.718          &    0.084     &    km~s$^{-1}$   \\
$K_2$    &   267.13         &    1.37       &    km~s$^{-1}$    \\
$V_0$    &   62.541          &    0.076     &    km~s$^{-1}$   \\
$q_{\rm sp} = M_2/M_1$ &   0.1300   &   0.0010   &                           \\
$A \sin i$  &  3.527        &     0.016     &   $R_\odot$        \\
$(M_1+M_2) \sin^3 i$ & 1.685  &  0.023  &  $M_\odot$    \\
$M_1 \sin^3 i$ & 1.491  &  0.020  &  $M_\odot$    \\
$M_2 \sin^3 i$ & 0.1938  &  0.0026  &  $M_\odot$    \\
\enddata

\tablecomments{
The period $P$ was assumed for the epoch 
of the observations, as described in the text.
The time of the deeper (transit) eclipse $HJD(pri)$ is from the
primary-component velocity curve. 
The radial-velocity amplitude of the secondary component was obtained from
measurements of the  profile centroid locations in 2D images, 
such as in Figures~\ref{night12}--\ref{night28}, with the
value of $V_0$ assumed to be the same as for the primary component.\\
The errors are formal least-squares errors and do not reflect 
systematic uncertainties in the data -- see the text in Section~\ref{sec}
for more details. 
}

\end{deluxetable}

%

\begin{deluxetable}{lccccl}

\tabletypesize{\footnotesize}                 
\tablewidth{0pt}
\tablecaption{Rotation of the $\epsilon\,$CrA components
\label{tab_RV}}

\tablehead{
\colhead{ } & \colhead{Phase} & \colhead{Observed} & \colhead{$L_1$} & 
    \colhead{$L_2$} &  \colhead{Observer sees} 
}
\startdata
Primary     &  0.25   &   147  &    178   &  192  & primary approaching      \\
                  &  0.5    &  143    &   173   &   184  & occultation of the secondary \\
                  &  0.75  &  147    &   178   &   192  & primary receding     \\
Secondary  & 0.25   &   89    &   74    &  114    & secondary receding   \\
                   &  0.0    &   66    &    64  &     75    & secondary transits the primary\\
                   & 0.75   &   57    &    74  &   114    &  secondary approaching   \\
\enddata

\tablecomments{
The values of the equatorial projected velocity $V \sin i$ are 
expressed in km~s$^{-1}$. They were determined by the single-star 
rotational-profile fits for the primary and as the profile half-widths at the base
for the secondary component (see the text). 
The observed values were determined to 
approximately $\pm3$ km~s$^{-1}$. 
Predictions of the solid-body rotation values are for the 
common equipotential surfaces passing through the 
Lagrangian $L_1$ and $L_2$ points.
The assumed sum of the observed orbital velocities used for the 
solid-body rotation estimates is $K_1+K_2 = 301.9 \pm1.5$  km~s$^{-1}$.
The geometry of the Roche model for $q = 0.13$ is from \citet{PK1964}.
}

\end{deluxetable}


\begin{deluxetable}{CC}

\tabletypesize{\footnotesize}                 
\tablewidth{0pt}
\tablecaption{Orbital velocities of the secondary component of $\epsilon\,$CrA 
\label{tab_sec}}

\tablehead{
 \colhead{$Time (Orb)$} & \colhead{Vel (km~s$^{-1}$)} 
}
\startdata
      0.7638   &   -211.41 \\
      0.7819   &  -207.11 \\
      0.7929   &  -192.50 \\
      0.8214   &  -178.75 \\
      0.8373   &  -161.56
\enddata

\tablecomments{
The first column gives the time in the $Orb$ units
(the time expressed in the binary orbital period;
see the text); the second column gives the 
radial velocity of the secondary component in km~s$^{-1}$, estimated
as a centroid of the secondary profile on the 2D images (see the text).\\
This table is available in the on-line version only. 
}

\end{deluxetable}

The orbital radial velocities of the secondary component were determined from 
the 2D images (Figures~\ref{night12}--\ref{night28}) 
for 184 epochs as centroids of the secondary-component profiles.
In addition to the orbital motion, we observed variations of 
the secondary profile in its width and structure, 
most likely because of the complex gas flows around
and over the surface of the secondary component.
The secondary profile shows more structure and is narrower after 
the secondary emerges from behind the primary ($\phi = 0.5$) than 
after the transit over the primary ($\phi = 0.0$) when the secondary recedes
from the observer. 

\begin{figure}[h]
\begin{center}
\includegraphics[width=12.5cm]{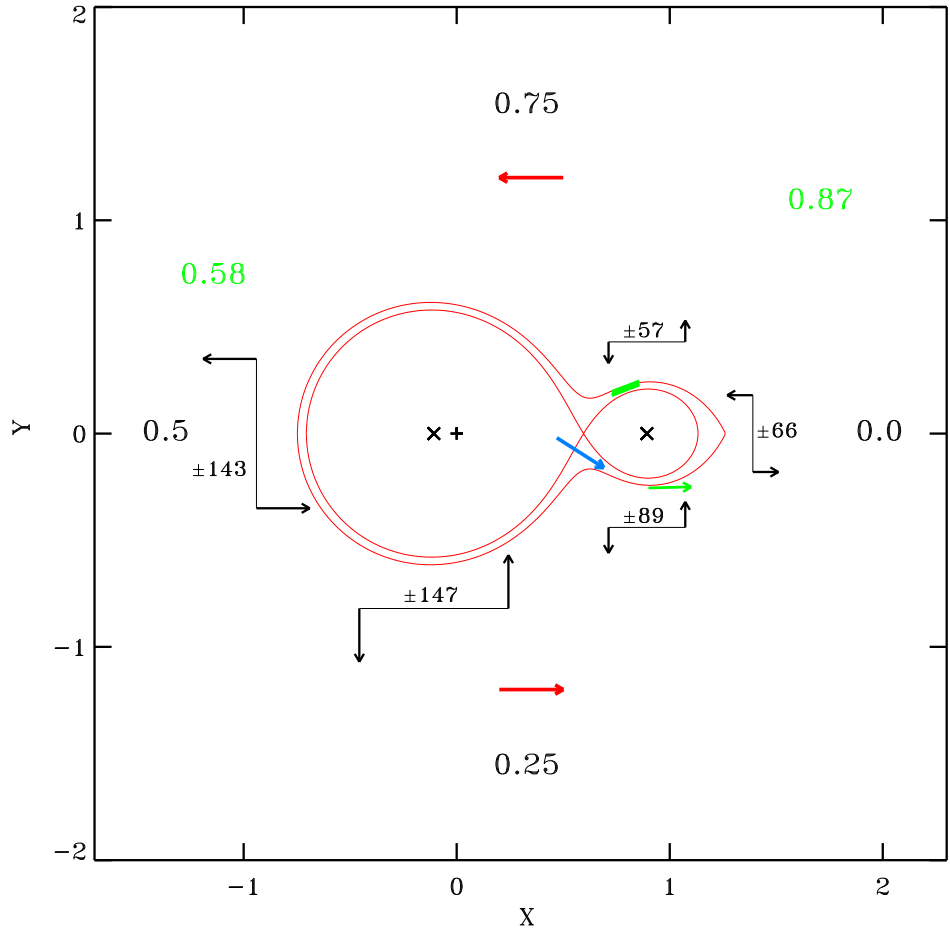}    
\caption{
\footnotesize 
A schematic view of the $\epsilon\,$CrA binary with the observed 
radial velocities shown as vectors. 
The unit of geometrical length in the orbital plane
is the separation between the mass centers; the velocity vectors are not to scale.
The  phase directions to the observer are written for four cardinal directions in black. 
The Roche model equipotentials for $q = 0.13$ are shown in red color; 
the red arrows mark its counter-clockwise rotation of the binary in the
figure frame. The individual rotation
velocities, such as the $V \sin i$ for the primary and the half-widths at
the base for the secondary are written in black as numbers preceded by the
$\pm$ sign. The identified sources of the secondary-component profile
perturbations are shown in green.
The green numbers for the phases correspond to the begin and end of the 
"wisp" visibility (see the text). The blue arrow shows the direction of the expected
flow originating in the primary component.
}
\label{Roche}
\end{center}
\end{figure}

The individual profiles of the secondary
do not appear to be dominated by rotation, as for the primary 
component, nor do they have shapes expected by the Lucy model. Thus,
we cannot directly identify their observed width with $V \sin i$; we can
however determine their observed widths from the relatively sharp edges 
at their BF base. The half-widths of the secondary profiles
vary gradually with the orbital phase. 
The widths can be best established at the phases of
the maximum velocity of approach and recession, in addition to the
direct radial-velocity imprint at the transit eclipse; the measurements 
at intermediate phases are more affected by the poor definition 
of the profile edges and could not be used. 
The observed half-widths range between 57 km~s$^{-1}$ at the
maximum approach of the  secondary ($\phi = 0.75$) 
and 89  km~s$^{-1}$ at its recession from the observer ($\phi = 0.25$); 
with the half-width at the transit eclipse ($\phi = 0.0$) of  66 km~s$^{-1}$
falling in between. 
The above numbers carry an uncertainty of about $\pm3$ km~s$^{-1}$.
The half-widths at the profile base,
interpreted as corresponding to $V \sin i$ are summarized in Table~\ref{tab_RV},
together with the half-widths expected for the rigidly-rotating Roche structure; 
Figure~\ref{Roche} gives a schematic distribution of
velocity vectors in the $\epsilon\,$CrA  system.

The centroid velocities of the secondary component profiles were determined
to $\pm3.5$ km~s$^{-1}$, as estimated from repeated measurements.
They follow the sine curve, as shown in Figure~\ref{fig_RV}.
The orbital parameters given in Table~\ref{tab_els} assume that 
the center-of-mass velocity, $V_0$, of the secondary is 
the same as for the primary component. While the uncertainty of $K_2$
of $\pm1.37$ km~s$^{-1}$ is small, compared with orbital solutions for other 
W~UMa binaries, 
the single-point error for the secondary, $\pm14.7$  km~s$^{-1}$ is large
and reflects systematic trends in the deviations within some of the individual orbits 
reaching $-30$ km~s$^{-1}$ to $+40$ km~s$^{-1}$. Since neither the Lucy model 
nor the simplified sine-curve fits describe such deviations, we
treat them as contributors to the random errors. As the result,  
the orbital parameters presented here should be treated as provisional
and subject to re-evaluation once a correct model becomes available. 
We note that a solution for the secondary with the mean systemic 
velocity $V_0$ treated as a free parameter differed in the mean velocity by 
$\Delta V_0 = -0.94  \pm1.77$ km~s$^{-1}$ from that for the primary component, 
indicating the overall level of systematic uncertainties in our 
data processing and velocity measurements. 

The quantity which should be least affected by the inadequacy of the 
adopted model should be the mass ratio, $q = K_1/K_2$. The light curve
synthesis simulations provide information on the mass ratio which is
independent of the spectroscopic result; however, it entirely depends 
on the adopted Lucy-model.  
The most recent, very accurate  photometric value of the mass
ratio, $q_{\rm ph}  = 0.1244 \pm0.0014$ by \citet{WR2011},
is only slightly smaller than our spectroscopic result, 
$q_{\rm sp} = 0.1300 \pm0.0010$; the latter is practically identical with
the result  by \citet{GD1993}, $q_{\rm sp} = 0.128\pm0.014$.
Apparently, the systematic difference, $q_{\rm ph} < q_{\rm sp}$,
discovered by \citet[Sec.5]{ND1991} is very weakly present 
in $\epsilon\,$CrA, only at the $4 \sigma$ level, and with error
estimates which may be systematically too optimistic\footnote{For a fair comparison,
all errors given here are the formal least-square estimates as is customarily
practiced for photometric analyses; however, systematic uncertainties 
do affect both techniques. This subject requires a separate careful study and
is not strictly related to the analysis of $\epsilon\,$CrA.}. 
Apparently, the similarity of $\epsilon\,$CrA to AW~UMa -- 
analyzed in the same way and using the same technique 
in Paper~I -- does not extend into the mass-ratio 
discrepancy.
This subject is discussed further in Section~\ref{comp}

\vfill

\subsection{Complications in the secondary component profile}
\label{stream}

The profiles of the secondary component, 
as shown in Figures~\ref{night12}--\ref{night28}, are fairly complicated
yet appear to be surprisingly stable, 
suggesting a binary-wide velocity  field in stationary conditions.  
However, it is not entirely obvious what is the cause of the details in the
profiles. Our Broadening Function information-extraction 
method is not sensitive to the rarefied gas:
It can only detect the presence of  a dense, high-temperature gas 
with similar characteristics to those of the photospheric layers of the 
primary component. If not projected against a photosphere,
the gas must be sufficiently dense to accumulate to the optical depth 
of at least order of unity along the line of sight to produce an
absorption feature detectable by the BF technique.
This condition may be fulfilled in localized places
close to the secondary component and possibly in the region between the 
binary components. The radial-velocity measurements give us
information about one velocity component of the 
gas motions only, not about localization of
the radiating gas. In addition to the eclipse shadowing effects,
a model or plausible assumptions
are the only possibilities to visualize the actual patterns of the gas
motions. 

The BF spike at the negative-velocity of the secondary 
component during its transit was described in  Section~\ref{sec} 
as defining the approaching side of that star
(see the lowest panel of Figure~\ref{fig_4ph}).
It is one of the most prominent manifestations of the
inter-binary matter, but it is visible only in a narrow range of phases
around $\phi = 0.0$, becoming the strongest at the exact alignment of the
stars. Its phase dependence is difficult to measure 
because the spike is small in the absolute sense 
with the integrated intensity at its maximum of 
less than one percent 
of that of the primary-component integrated profile. 
If  strengthening of the spectral lines was coincident with a continuum 
increase, such a faint feature would be detectable in broad-band photometry,
possibly combining with other small variations during the eclipse branch. 
In the case of the current program, 
it was possible to detect it spectroscopically mostly thanks 
to the linear properties of the BF technique and after 
subtraction of the dominant primary-component profile.
Its width at the base is $25 \pm2$ km~s$^{-1}$,
as  estimated from repeated transit events.
We think that the spike is produced by the outer layers of
a stream of matter flowing away from the 
primary component and sliding along the surface of the secondary,
directly towards the observer.  The outer layers of the flow
attain a sufficient optical depth when visible along and into their length;
they must be relatively slim and long to produce a strong 
dependence on the angle. 

\begin{figure}[h]
\begin{center}
\includegraphics[width=11.5cm]{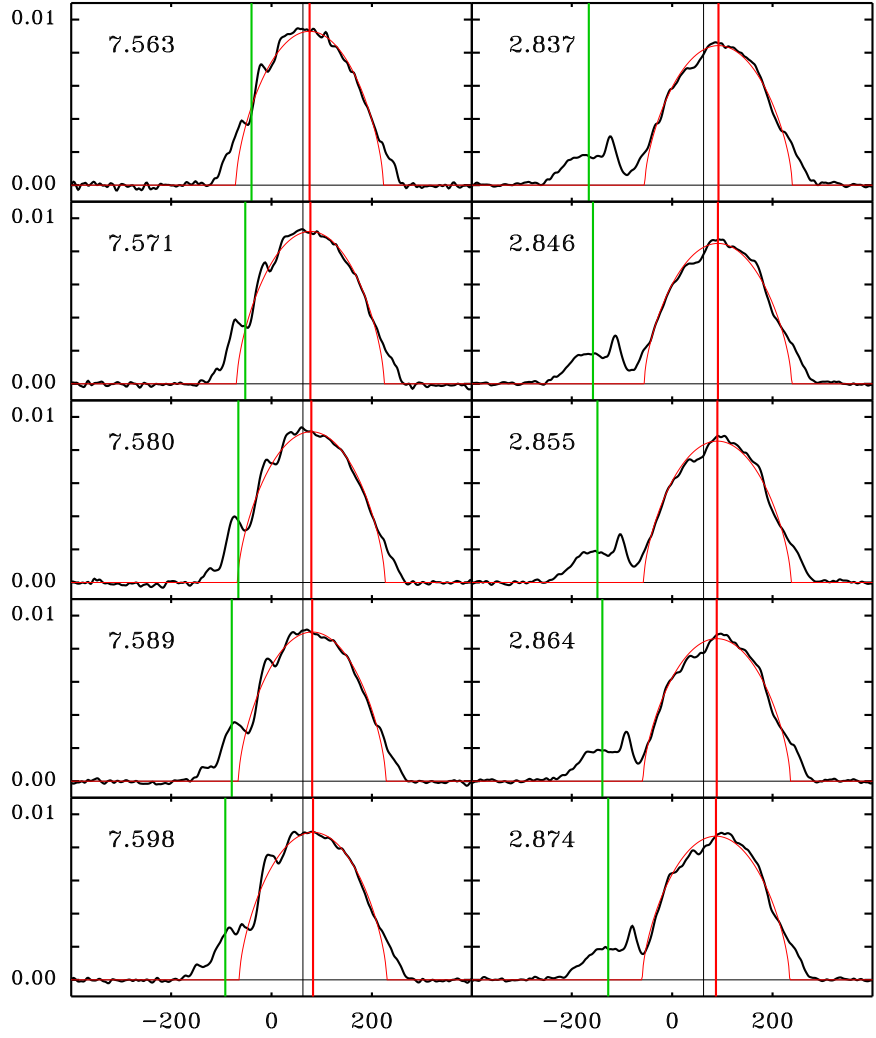}    
\caption{
\footnotesize 
The Broadening Functions illustrate the secondary-profile complication
visible as a radial-velocity feature called a ``wisp'' in the text. 
The wisp is interpreted  as a radial-velocity representation of a stationary
area of the enhanced absorption spectrum. 
The left column shows a sequence for the orbital phase
around $\phi \simeq 0.58$ when the wisp appears after 
the secondary occultation while the right column shows the orbital phases
close to $\phi \simeq 0.86$ 
when the projection of the secondary component starts approaching its 
transit over the primary component. 
The vertical red and green straight lines give the orbital velocities of the 
mass centers of the primary
and secondary components determined by the sine-function fits, while 
the black line gives the velocity of the binary center. 
The rotational profile of the primary for $V \sin i = 147$ km~s$^{-1}$
is shown by a thin red line. The $Orb$ number is given in the left upper corner of
each panel; its fractional part is the orbital phase. 
}
\label{fig_stream}
\end{center}
\end{figure}

The radial-velocity profile of the secondary component appears to be more 
complicated within the orbital phases when the secondary approaches the 
observer (the phase interval $0.5 < \phi < 1.0$). 
The most prominent feature visible in the radial velocity images in 
Figures~\ref{night12}--\ref{night28}
is called here a ``wisp'' (see also the third panel of  Figure~\ref{fig_4ph}
and Figure~\ref{fig_stream}). The wisp emerges from
behind the primary component around the orbital phase $\phi \simeq 0.58$
and remains visible to  $\phi \simeq 0.87$. The phases of 
the wisp appearance and disappearance 
are the only positional indications on the location where the wisp is formed,
but these limiting phases are difficult to determine; 
they are  coincident with the intervals when 
the radial-velocity measurements of the primary component required
rejection of several discrepant 
measurements from the final velocity curve (Figure~\ref{fig_RV}).

\begin{figure}[h]
\begin{center}
\includegraphics[width=8.0cm]{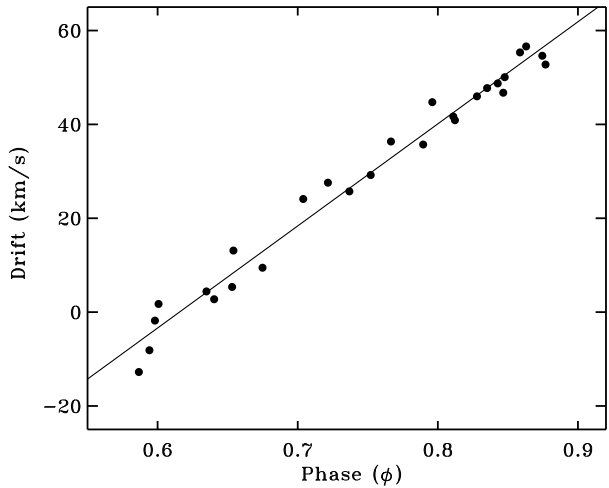}    
\caption{
\footnotesize 
Radial velocities of the  ``wisp'' visible in the 2D 
images in Figures~\ref{night12}--\ref{night28} within the orbital-phase range
when the secondary component approaches the observer. The Drift velocity is
expressed relative to the orbital velocity of the secondary component. 
}
\label{fig_wisp}
\end{center}
\end{figure}

The wisp has very different properties from the brief BF spike 
at the binary-component alignment producing the transit eclipse: 
It is not rapidly changing its intensity with
the orbital phase, instead, it shows a systematic drift in radial velocities 
within its visibility phase range,
$0.58 < \phi < 0.87$ (the best visible in Figure~\ref{night13}; see also
Figure~\ref{fig_stream}). 
It stays confined to the middle of the secondary-component profile 
suggesting a feature possibly anchored on the stellar disk and
apparently confined to one hemisphere since it
does not reappear on the positive-velocity side of the system when
the secondary component recedes from the observer ($\phi > 0.0$). 
The wisp tends to be wider soon after its emergence from
behind the primary component, with the full
width at the base of about 35--40  km~s$^{-1}$, then it 
narrows down to 15--20  km~s$^{-1}$ before its disappearance. 
Its migration in the radial velocities relative to the center of mass
of the secondary component can be approximated by
a linear dependence on the orbital phase, 
${\rm Drift} = 29.24(\pm0.64) + 217.50(\pm6.53) \times (\phi - 0.75)$  km~s$^{-1}$
km~s$^{-1}$
(Figure~\ref{fig_wisp}), where the single-measurement error of the relation, 
3.3  km~s$^{-1}$, is mostly due to measurements uncertainties.

Assuming the projected rotational velocity of the secondary component of
$-66$ km~s$^{-1}$ and the orbit-synchronous rotation,
the constant intensity and the apparent 
radial-velocity drift of the wisp can be explained by
a spatially-stationary patch located on this component and visible for
half of the orbital revolution. 
This patch of an increased strength of the spectral lines would be then
located on the positive Y-axis side in Figure~\ref{Roche},
at about $60 - 65$ degrees away from the line joining the stars. 
This location would  explain why the patch is invisible
within the orbital phase range $0 < \phi < 0.5$, as well as the   
early appearance after the mid-occultation at  $\phi = 0.58$
and an even-earlier disappearance before the mid-transit 
at $\phi = 0.87$. These numbers assume a 
moderately low orbital inclination angle of $i \simeq 70$ degrees 
producing a grazing totality eclipse.
The wisp, as a positive BF feature with the strength of about $1.5-2$\% 
of the primary component integrated BF intensity, is produced by a localized
enhancement of the absorption line spectrum. Although it is so weak,
its strength is not negligible relative to the secondary-component 
where it contributes up to 20\% of the integrated BF intensity. 

\begin{figure}[h]
\begin{center}
\includegraphics[width=9.5cm]{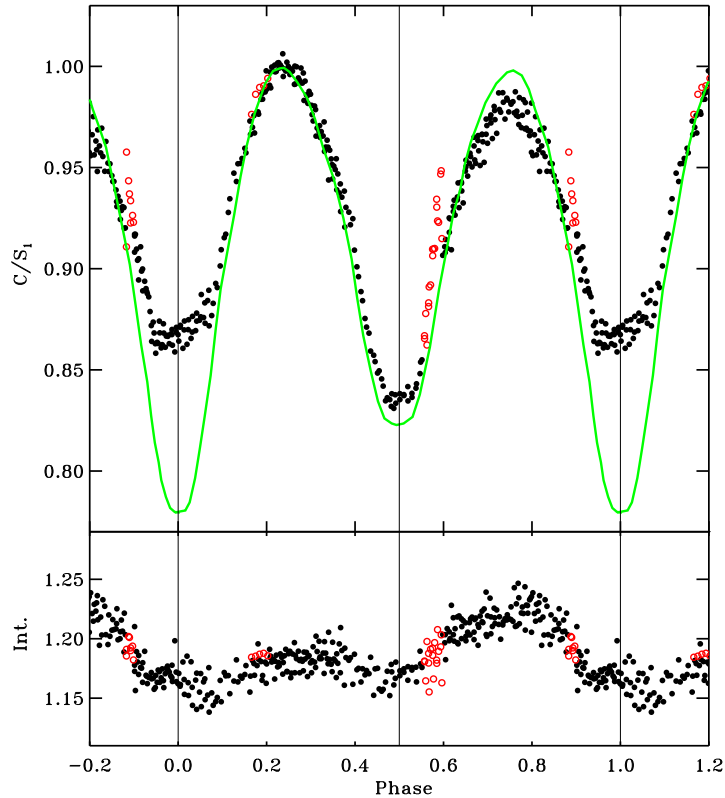}    
\caption{
\footnotesize  
The ``light curve'' of $\epsilon\,$CrA restored from 
the BF spectral data (upper panel) and
the phase variations of the integrated equivalent width (IEW) of the spectral lines; 
see the text for details.
The red, open circles mark the data not used for the radial-velocity
orbit determination. 
The green line shows the visual-band light curve obtained by 
\citet{Knip1967} after conversion to flux units. 
}
\label{LC}
\end{center}
\end{figure}

The input spectra were individually normalized (to the unit level at the observed
continuum) so that without simultaneous 
photometric observations we can only state that the spectral lines 
become stronger during the visibility of the wisp. But the mechanism is
not known as the lack of an independent information on the continuum 
precludes a proper interpretation of the wisp. We see two possible
explanation of the enhanced spectrum producing the wisp:
\begin{enumerate}
\item The patch of the stronger lines is produced by a 
localized piling up of the dense gas. Normally, added material
on top of the atmosphere would simply shift the location of the
surface corresponding to the optical-depth of one. But, since
we do not know the exact geometry of the gas circulation 
in the $\epsilon\,$CrA system, this may be a viable solution.
The broad-band photometric effect of such
a gas concentration is expected to be
very small at a level of $\le 0.1$\%
since the wisp corresponds to a $\le 2$\% modification of
the spectrum whose integrated equivalent width of all spectral lines in our 
spectral window is about 5.5\% relative to the continuum level.
\item A spot of a depressed effective temperature. 
The locally lower temperature would lead to a decrease of the star 
brightness and to a richer and stronger absorption-line spectrum.
Our spectra were normalized 
so that information on the stellar brightness was lost. 
This matter could not be resolved without broad-band data, but  
we have indirect indications (see the next Section~\ref{light}
and Figure~\ref{LC}) that the light-curve maximum after the
secondary eclipse ($\phi \simeq 0.75$) was by about 2\% lower 
than the first maximum. This would
suggest that the wisp-causing dark spot may have indeed influenced 
the continuum level. Such a spot would be most likely 
very different from magnetic star spots because the effective temperature 
drop must be rather moderate. 
Usually magnetic spots are dark and it is hard to detect 
their spectral contribution, whereas here the line-spectrum enhancement 
would  over-compensate for the brightness decrease. 
\end{enumerate}

\subsection{The spectroscopic ``light curve'' }
\label{light}

We demonstrated for the case of AW~UMa (Paper~I)
that, with some assumptions, the spectroscopic 
results can give a direct link to the photometric phase variability of this binary.  
We utilized (1)~the 
BF normalization of the integrated equivalent width (IEW) of all spectral 
lines in the spectral window to the IEW of the model 
spectrum,  and -- building upon the 
observed approximate constancy of the effective temperature with the orbital 
phase, (2)~we made a strong assumption that the spectral lines of the 
primary component remain constant with the orbital phase. 
The results for $\epsilon\,$CrA are shown in Figure~\ref{LC}, while additional
explanations of the two assumptions  are as follows:
\begin{enumerate}
\item The BF's for AW~UMa and $\epsilon\,$CrA
were determined using the same model atmosphere, F2V for the solar
abundance. The IEW in  AW~UMa showed only a very small IEW 
phase variation of  about 2\% around the value of 1.10 (Figure~2, Paper~I).
This agreed with the approximate effective-temperature constancy and
could be interpreted as a slightly later spectral type of AW~UMa than 
of the template. In $\epsilon\,$CrA, the IEW is larger, 
with the mean value of 1.18 (Figure~\ref{LC}), in agreement with suggestions
of an even later spectral type of the binary. 
The IEW shows about 5\% departures from the mean value 
reaching 1.23 at the orbital phases around $\phi \simeq 0.75$.
We note that these are the phases when the
secondary component approaches the observer and when 
the wisp-producing feature (Section~\ref{stream}) 
could influence the continuum spectrum; apparently, the 
feature produces stronger absorption lines as if the local effective
temperature was lower. 
The IEW for $\epsilon\,$CrA is not as invariable with phase 
as that for AW~UMa; the difference between 
the binaries may possibly relate to their  different orbital inclination angles
and the different visibility of their orbital planes where -- according to the model
of St\c{e}pie\'{n} (Section~\ref{new}) -- 
the energy is transported between the binary components. 
\item The assumption (2) states that the primary 
component does not change with the orbital phase
and remains a rapidly rotating star describable by a rotational profile;  
 it is oblivious to the existence of the secondary component. 
In other words: Whatever happens in the system in terms of the line
producing photospheric area is due to the secondary and the matter around it.
The assumption is made simply for convenience; its correctness
is verifiable by the result.  
\end{enumerate}

The ``spectroscopic light curve'' is calculated using 
the maximum value ($S_1$)  of the primary component
rotational-profile to form
a phase dependence of its inverse $C/S_1$ (see the discussion in
Paper~I, Sec.~3). 
It turns out that such dependence is a very reasonable approximation 
to the photometric light curve, as shown in the upper panel of 
Figure~\ref{LC}, confirming our assumptions. 
In Paper~I, we used for $C$ the normalized integral of the
rotational profile, but here we simply use an arbitrary constant adjusted
to produce a light curve normalized to unity at the maximum light. 
Such a light curve is 
correctly phased and can be compared with the calibrated light curves of 
\citet{Knip1967} (shown in Figure~\ref{LC}), \citet{Tapia1969}
and \citet{Hernan1972}. Only the primary (transit) eclipse
comes out too shallow, but this is expected since 
the assumption of no apparent change to the primary component
profile  breaks down during these phases.  
 
In very simple terms, the ``spectroscopic light curve'' 
derived in the above way  expresses the amount of the spectral-line
producing surface projected onto the sky. 
Figure~\ref{LC} tells us that this
amount is very close to what can be calculated 
using the Lucy model. Thus, such a spectroscopically-derived 
light curve does carry the same information about the binary -- but
admittedly in a more convoluted way -- as a standard photometric light  
curve. 

The successful application of our approach
may turn out to be restricted to the low mass-ratio systems 
such as our two targets, AW~UMa and $\epsilon\,$CrA,
where the primary components strongly dominate in the energy balance.
However, thanks to the steep mass-luminosity relation in the lower 
Main Sequence, the domination of the primary component
may extend into moderate mass ratios. The binary parameter range 
for the above approach remains to be determined. 

\bigskip

\section{Conclusions}
\label{mod}

\subsection{$\epsilon\,$CrA and AW~UMa}
\label{comp}

The new, spectral time-monitoring observations of $\epsilon\,$CrA 
show the binary as a twin to AW~UMa analyzed in Paper~I. The observations
were more extensive (3.1 times more total exposure time)
and more accurate (the observational $S/N$ improved about twice)
so that the results carry more weight in a  discussion of both binaries.  
$\epsilon\,$CrA appears to be slightly more massive, as 
expected for a longer orbital period, but the spectral types 
are practically identical, although $\epsilon\,$CrA may
have more evolved cores judging by its larger luminosity and the slightly
lower effective temperature. 
The two binaries are very similar in having  small 
mass ratios ($q_{\rm sp} = 0.10$ and 0.13). Most importantly,
both show an identical disagreement 
with the assumptions of the Lucy model by having 
an apparently stationary velocity field in the rotating frame of the
binary. The gas flows are best visible within the secondary-component
profile and in the volume between the components.
We did not see any obvious changes in the flow velocity patterns 
 over 27 orbital periods so  $\epsilon\,$CrA may have 
more stable internal velocities than AW~UMa; the latter, however,
was not observed equally extensively as $\epsilon\,$CrA. 
In $\epsilon\,$CrA, 
a ``wisp'' (Section~\ref{stream}) in the 2D time-sequence images of the
surface velocity motions, associated with the secondary component was seen
to be stable over the duration of the observations. 
Similar, but less well-defined velocity features were seen in AW~UMa.  
Such features are most probably produced
by  gas streams seen along their longer dimensions through accumulated
optical depth or may come from isolated areas of enhanced spectral-line
strength of currently unexplained properties.  

The general stability of both binaries is confirmed by the moderate
systematic period variations, with $dP/dt$ at a level of 
a few $\times 10^{-7}$ day/year. The period
changes in these two binaries are of opposite sign, hence they are most 
likely unrelated to their secular evolution. They may be explainable 
by a small imbalance in the net mass transfer between the binary
components or by external perturbations; we recall that W~UMa binaries
practically always exist in triple systems,
 see  \citet{Toko2006} and \citet{Rci2007}. 
Since large amounts of matter carrying radiation energy 
appear to circulate between the components (see the next 
Subsection~\ref{new}), 
the flows must be well balanced in terms of the net amount 
of the matter transferred between the stars. 
 
The primary in both binaries
is a rapidly-rotating early F-type star with the projected 
equatorial rotational velocity $V \sin i$ at about 80\% of the
rate synchronous with the orbital motion (cf.\ Section~\ref{pri}).
The BF profile appears symmetric and well represented by the standard
rotational profile. The slower-than-expected rotation suggested for the higher 
latitudes of the primary may partially result from the 
anti-cyclonic gas motion (in the counter-rotation direction)
seen in the mass-losing, hydrodynamic models of \citet{Oka2002}. 
Gas velocities in such a high-pressure cell are expected to link
the high stellar latitudes of the primary to the outflow from
the Lagrangian $L_1$ point. Only a modest reduction of the observed 
rotational velocity is expected however, perhaps 
by a value of the order of sonic or subsonic photospheric velocity which
for the spectral types of both stars is $\simeq 7$  km~s$^{-1}$.
The pedestal at the base of the primary rotational profile 
with the additional  broadening by 25--35 km~s$^{-1}$ 
would bring the observed $V \sin i$ 
into an approximate agreement with the synchronous
rate with the orbit. The pedestal contributes moderately to the
total depth of the spectral lines, typically at a level 0.06 to 0.10 
of the integrated rotational profile of the primary.  The
pedestal likely reflects the presence of the matter in the 
orbital plane, returning to the primary after its orbit around the secondary
component.

The secondary components of W~UMa-type binaries are important for
our understanding of the unresolved secret of how these binaries
function as stable structures. The secondaries of both,
$\epsilon\,$CrA (Section~\ref{sec}) and AW~UMa (Paper~I), look similar in that
they show orbital-phase changes which are repeatable and 
stable in time, and appear to be related to the phase-dependent
visibility of an otherwise stationary velocity field. 
We identified two features in the somewhat complex radial velocity
picture:
(1)~A stream directed toward the observer over the secondary surface, 
accumulating to sufficient 
optical line depth within a narrow cone of visibility;
(2)~A region where the absorption line spectrum is for some reason
stronger than in the surrounding; it can be a region of a  local,
geometrically complex piling up of the gas possibly through
collision of a direct and circum-secondary flows from the primary, or an area
of a slightly lower effective temperature with enhanced absorption 
line spectrum. 
These  features may be the sought-after surface imprints of a deeper 
energy transport from the primary to the secondary component. 
In spite of the complex radial-velocity picture, 
the amount of the visible, line-producing photosphere of
the secondary and of the matter between the stars approximately
agrees with that predicted by the Roche-model based model 
of \citet{Lucy1968b}. 

\bigskip

\subsection{A new model for W~UMa binaries?}
\label{new}

Our observational results for AW~UMa and $\epsilon\,$CrA
relate to two small mass-ratio W~UMa-type systems. Despite their
small mass ratios, both binaries have been always
considered genuine members of and representatives for 
the whole class of these objects. We assume that our results indeed
apply to the whole class of the W~UMa-type binaries. 

The available radial-velocity data for AW~UMa and $\epsilon\,$CrA
show that the main assumption of the \citet{Lucy1968a,Lucy1968b} model,
the solid-body rotation of the components as a prerequisite 
for the definition of the Roche-model equipotentials, is not confirmed.
A complex system of velocities in the rotating system of coordinates
is particularly well visible within the phase-dependent radial-velocity
profiles of the secondary component. The velocities are moderate, of
the order of or less than the rotational velocity of the secondary
component while their observed ranges are of the order of the local sound
speed. But their spatial pattern is of great importance as we see circulation
fully confined to the secondary component, similar to that expected on 
a non-contact binary component. The impact of such well organized
velocity systems on the assumed W~UMa-binary model and on
interpretation of light-curve observations is currently unknown and
should be investigated. 

The  solid-body assumption at the time of the Lucy model creation
was simply the easiest to make, although no mechanism had been 
offered why the components would rotate in such a way.
This assumption made the Lucy model particularly easy 
to use as a light-curve modeling tool so that 
 many light-curve solutions have 
been made utilizing the model's simplicity. 
Most of these solutions produced photometric mass ratio,
$q_{\rm ph}$, determinations utilizing the one-to-one mapping between
the mass ratio and the relative dimensions of the components,
first discussed by \citet{MD1972}. While this relationship
has been particularly useful for light-curve solutions, 
it appears that it may require a correction or modification. 
This had been  signalled by \citet{GD1993}, 
was strongly confirmed for  AW~UMa and is now indicated for
$\epsilon\,$CrA. 
While the discrepancy between the spectroscopically determined
$q_{\rm sp}$ and photometrically $q_{\rm ph}$ for $\epsilon\,$CrA 
is indicated at the level of $4 \times \sigma$ (with possibly over-optimistic
error estimates), for  AW~UMa the relative
discrepancy $(q_{\rm sp} - q_{\rm ph})/q_{\rm sp}$  
reached the unbelievable relative value of $\ge 10$\%! 
This suggests that either the spectroscopic or photometric 
determinations of $q$ may be biased. We point out that unrecognized,
but omnipresent companions tend to push photometric solutions towards
smaller mass ratios and/or smaller 
orbital inclinations by a photometric-scale change; 
no equally simple biasing mechanism exists for spectroscopic
mass-ratio determinations executed at an adequate resolution. 
Whatever the reason, the continuing routine use of the Roche-lobe based
photometric model to determine the values of $q_{\rm ph}$ 
may give systematically wrong individual results, biasing 
statistics of the mass ratio, masses and orbital inclination angles. 

The \citet{Lucy1968a,Lucy1968b} model, when it was developed
more than a half-century ago, offered the first physically acceptable 
description of W~UMa-type binaries. It generated an extensive interest,
but also considerable controversy after its implied properties were 
studied in more detail. The main difficulties were related to the 
internal structure and effective-temperature equalization processes. 
Further research led to the consideration of the thermal cycles consisting of
semi-detached and contact configurations 
\citep{Lucy1976,Flan1976}, but the problems did not disappear, but related 
to the structure during a part of the thermal cycle,
after re-establishment of the contact.  
The discussion culminated in detailed studies of the energy transport 
in contact by \cite{Webb1976,Webb1977} and a heated discussion
of the concept of a ``contact discontinuity'' 
\citep{Shu1976,Hazle1978,Papa1979,Shu1979}.
The ``discontinuity'' concept was important as it envisaged the secondary 
component fully enshrouded in the matter of the primary component.
This assumption solved the equality of the effective temperatures while the 
rest of the observables, particularly 
the solid-body-rotating equipotential structure, remained as
in the original Lucy model. The end to further discussion and similar
first-principle considerations came from the realization that -- with only
photometric data -- there exist no 
observational verification methods for this complex problem.
The model concept of \citet{Step2009} returned to the idea of
the gas of the primary-component surrounding the secondary, but
now not as a static equipotential structure, but rather as a stationary
flow of the optically-thick matter confined to the equatorial 
regions and eventually returning to the primary. 
In its motion, the flow is controlled by
the gravitational pull to the orbital plane and the dispersing 
tendency of the Coriolis force in the plane. 
The St\c{e}pie\'{n} study specifically addressed the energy transport
requirements and did not offer any predictions
on the observational consequences of the flow, making his ideas 
 less directly applicable at the present stage than the Lucy model.

Our detection of the organized gas flows in AW~UMa and $\epsilon\,$CrA
supports the energy-transport model of \citet{Step2009} 
and the pioneering hydrodynamic calculations of \citet{Oka2002}.  
The essential component of a new vision is a stationary flow  
that dominates the thermal equilibrium of the binary and which
starts and ends on the primary component. 
The flow  submerges the equatorial regions of the secondary in 
optically thick gas. It leaves the primary at  high latitudes 
to slide down to the low latitudes of the $L_1$-point region where it starts its 
orbit around the secondary component, first curving to one side of the
binary (the negative Y-axis side in Figure~\ref{Roche}), and then returning
to the primary at its low latitudes.
When all the matter returns to the primary and there is 
no net mass-transfer between the components, then 
no orbital period changes are expected to be observable.  
The stream at its return may lead to formation of the
perturbed region identified between the stars in $\epsilon\,$CrA;
it may also excite the standing atmospheric waves 
visible as the ripples in AW~UMa. We see only the outermost layers 
of the moving matter, although the process may involve large amounts 
of matter carrying substantial energy. Generally, only 
moderate observable effects are expected, such as  
those we detected in AW~UMa and $\epsilon\,$CrA, 
but some details are not easily explainable,
e.g.\ the phase-dependent velocity extent of the secondary 
component. 

Two directions can be followed for further research to develop 
modifications of the current W~UMa-binary model: 
The theoretical one specifically addressing the detailed
properties of the St\c{e}pie\'{n} energy-transfer
model, perhaps following the directions
established by the pioneering hydrodynamical models
of \citet{Oka2002} and the observational one based on high-resolution 
time-sequence spectroscopy. Both may present considerable difficulties.
The former would require extensive use of super-computers in 
a simulation incorporating complex physical phenomena, while the
observations would be difficult to arrange 
as they require relatively large amounts of time on moderate- or 
large-size telescopes equipped with high-efficiency, high-resolution 
spectrographs. 
As always, the observations would provide the necessary guidance,
but are harder to generalize than theoretical models. 
The result of the effort is expected to lead to a paradigm shift from 
an elegant picture of a  static, solid-body, Roche-model structure 
with the inexplicable equality of 
temperature over its surface, to a more complex vision of a 
gaseous binary approximately agreeing with the Roche model,
with the primary component engulfing the secondary component 
in its equatorial regions by an optically thick, stationary flow carrying
the temperature-equalizing energy. 

\bigskip

\begin{acknowledgements}
We appreciate the support of the CHIRON team, particularly 
Mr.\ Leonardo Paredes, and Drs.\ Rodrigo Hinojosa and Todd Henry for 
assistance in securing the observations via the SMARTS consortium
at the Cerro Tololo Observatory. The observations
have been supported by the grant from by the Natural Sciences
 and Engineering Research Council of Canada.
Prof.\ J.\ Kreiner and Dr.\ B.\ Zakrzewski are thanked for providing the 
eclipse ephemeris data.
Dr.\ Joel Eaton is thanked for a promise to take care of the paper and its
data during a particularly difficult time for the author. 
Special thanks are due 
to Prof.\ Kazik St\c{e}pie\'{n} for his constructive review of an 
early version of the manuscript, several discussions 
and a number of useful suggestions. 

This work has made use of data from the European Space Agency (ESA) mission
{\it Gaia} (\url{https://www.cosmos.esa.int/gaia}), processed by the {\it Gaia}
Data Processing and Analysis Consortium (DPAC,
\url{https://www.cosmos.esa.int/web/gaia/dpac/consortium}). Funding for the DPAC
has been provided by national institutions, in particular the institutions
participating in the {\it Gaia} Multilateral Agreement.
The research made also use of the SIMBAD database, 
operated at the CDS, Strasbourg, France and of NASA's 
Astrophysics Data System (ADS).
\end{acknowledgements}

\newpage

\newpage

\end{document}